\documentclass[12pt,a4paper]{article}
\usepackage[dvips]{graphicx,color}
\newcommand{\be}{\begin{equation}}
\newcommand{\ee}{\end{equation}}
\newcommand{\ba}{\begin{eqnarray}}
\newcommand{\ea}{\end{eqnarray}}
\newcommand{\baa}{\begin{eqnarray*}}
\newcommand{\eaa}{\end{eqnarray*}}
\newcommand{\lab}[1]{\label{#1}}

\newcommand{\dis}{\displaystyle}
\newcommand{\biq}{\mbox{\boldmath $q$}}
\newcommand{\bip}{\mbox{\boldmath $p$}}
\newcommand{\bif}{\mbox{\boldmath $f$}}

\begin{document}
{\pagestyle{empty}
\vskip 1.5cm

{\renewcommand{\thefootnote}{\fnsymbol{footnote}}
\centerline{\large \bf Generalized-Ensemble Algorithms for
Molecular Simulations of Biopolymers}
}
\vskip 3.0cm
 
\centerline{Ayori Mitsutake,$^{a,b}$
Yuji Sugita,$^{a,b}$
and Yuko Okamoto$^{a,b,}$\footnote{\ \ {\it Correspondence to:}
Y. Okamoto, 
Department of 
Theoretical Studies, Institute for Molecular Science,
Okazaki, Aichi 444-8585, Japan \ \ 
e-mail: okamotoy@ims.ac.jp}}
\vskip 1.5cm
\centerline{$^a${\it Department of Theoretical Studies}}
\centerline{{\it Institute for Molecular Science}}
\centerline{{\it Okazaki, Aichi 444-8585, Japan}}
\centerline{and}
\centerline{$^b${\it Department of Functional Molecular Science}}
\centerline{{\it The Graduate University for Advanced Studies}}
\centerline{{\it Okazaki, Aichi 444-8585, Japan}}

\vskip 1.5cm

\centerline{Submitted to {\it Biopolymers (Peptide Science)}}

\vskip 0.5cm

\centerline{{\it {\bf Keywords:} protein folding; 
generalized-ensemble algorithm; multicanonical algorithm;}}
\centerline{{\it simulated tempering; replica-exchange method; parallel tempering}}

\medbreak
\vskip 1.0cm
 
\centerline{\bf ABSTRACT}
\vskip 0.3cm

In complex systems with many degrees of freedom such as
peptides and proteins there exist a huge number of 
local-minimum-energy states.  Conventional simulations 
in the canonical
ensemble are of little use, because they tend to get trapped in
states of these energy local minima.
A simulation in generalized ensemble performs a random walk
in potential energy space and can overcome this difficulty.
From only one simulation run, one can
obtain canonical-ensemble averages of physical quantities as
functions of temperature
by the single-histogram and/or multiple-histogram reweighting techniques.
In this article we review uses of the generalized-ensemble algorithms.
Three well-known methods,
multicanonical algorithm, simulated tempering, and 
replica-exchange method, are described first.
Both Monte Carlo and molecular dynamics versions of the 
algorithms are given.
We then present three new generalized-ensemble algorithms
which combine the merits of the above methods.
The effectiveness of 
the methods for molecular simulations in the
protein folding problem 
is tested with short peptide systems.
 
}
\section{INTRODUCTION}
  
Despite the great advancement of
computer technology in the past decades, simulations of 
complex systems such as spin glasses and 
biopolymers are still greatly hampered by the multiple-minima
problem.
It is very difficult to obtain accurate canonical distributions
at low temperatures by conventional Monte Carlo (MC) and
molecular dynamics (MD) methods.
This is because simulations at low temperatures tend to get
trapped in one of huge number of local-minimum-energy states.
The results thus will depend strongly on the initial conditions.
One way to overcome this multiple-minima
problem is to perform a simulation in a {\it generalized ensemble} where
each state is weighted by a non-Boltzmann probability
weight factor so that
a random walk in potential energy space may be realized.
The random walk allows the simulation to escape from any
energy barrier and to sample much wider phase space than
by conventional methods.
Monitoring the energy in a single simulation run, one can
obtain not only
the global-minimum-energy state but also canonical ensemble
averages as functions of temperature by the single-histogram \cite{FS1}
and/or multiple-histogram \cite{FS2,WHAM} reweighting techniques
(an extension of the multiple-histogram method is referred to as
{\it Weighted Histogram Analysis Method} \cite{WHAM}).

One of the most well-known generalized-ensemble methods is perhaps
{\it multicanonical algorithm} (MUCA) \cite{MUCA1,MUCA2}
(for a recent review, see Ref.~\cite{MUCArev}).
(The method is also referred to as {\it entropic sampling}
\cite{Lee} and {\it adaptive umbrella sampling} \cite{BK}.
The mathematical equivalence of multicanonical algorithm and entropic
sampling has been given in Ref.~\cite{BHO}.)
MUCA and its generalizations have been applied to
spin glass systems
(see, e.g., Refs.~\cite{MUCA3}--\cite{MUCA6}).
MUCA was also introduced to the molecular simulation field
\cite{HO}
(for previous reviews of generalized-ensemble approach in the
protein folding problem, see, e.g.,
Refs.~\cite{RevO}--\cite{RevHO2}).
Since then MUCA has been extensively
used in many applications in
protein and related systems \cite{HO94}--\cite{MO3}.
Molecular dynamics version of MUCA
has also been developed \cite{HOE96,NNK} (see also
Refs.~\cite{Muna,HOE96} for Langevin dynamics version).
Moreover, multidimensional (or multicomponent) extensions of MUCA
can be found in Refs.~\cite{KPV,HNSKN,ICK}.
  
While a simulation in multicanonical ensemble performs a free
1D random walk in energy space, that in
{\it simulated tempering} (ST) \cite{ST1,ST2} 
(the method is also referred to as the
{\it method of expanded ensemble} \cite{ST1})
performs a free random walk in temperature space
(for a review, see, e.g., Ref.~\cite{STrev}).
This random walk, in turn,
induces a random walk in potential energy space and
allows the simulation to escape from
states of energy local minima.
ST has also been applied to protein folding
problem \cite{IRB1}--\cite{IRB2}.

A third generalized-ensemble algorithm that is related
to MUCA is {\it 1/k-sampling} \cite{HS}.
A simulation in 1/k-sampling
performs a free random walk in 
entropy space, which, in turn,
induces a random walk in potential energy space.
The relation among the above three generalized-ensemble
algorithms was discussed and the effectiveness of the
three methods in protein
folding problem was compared
\cite{HO96b}.  

The generalized-ensemble method
is powerful, but in the above three methods the probability
weight factors are not {\it a priori} known and have to be
determined by iterations of short trial simulations.
This process can be non-trivial and very tedius for
complex systems with many local-minimum-energy states.
Therefore, there have been attempts to accelerate
the convergence of the iterative process for MUCA
\cite{MUCA3,KPV,SmBr,H97c,MUCAW,BK}
(see also Ref.~\cite{MUCArev}).
  
A new generalized-ensemble algorithm that is based on
the weight factor of Tsallis statistical mechanics \cite{Tsa}
was recently developed with the hope of overcoming this difficulty
\cite{HO96d,HEO98},
and the method was applied to a peptide folding problem
\cite{HMO97,HOO}.  A similar but slightly different formulation 
is given in Ref.~\cite{Str2}.  See also Ref.~\cite{Muna2} for
a combination of Tsallis statistics with simulated tempering.
(Optimization problems were also addressed by 
simulated annealing 
algorithms \cite{SA} based on the Tsallis weight in
Refs.~\cite{STsal}--\cite{H97b}.  For reviews of molecular simulations
based on Tsallis statistics, see, e.g., 
Refs.~\cite{BeSt}--\cite{RevHO3}.)
In this generalized ensemble the weight factor is known,
once the value of the global-minimum energy is given \cite{HO96d}.
The advantage of this ensemble is that it greatly simplifies the
determination of the weight factor.
However, the estimation of
the global-minimum energy can still be very difficult.

In the {\it replica-exchange method} (REM)
\cite{RE1}--\cite{RE2}, the difficulty of weight factor
determination is greatly alleviated.  (A similar method was
independently developed earlier in Ref.~\cite{RE3}.
REM is also referred to as {\it replica Monte Carlo method} \cite{RE3},
{\it multiple Markov chain method} \cite{RE4},
and {\it parallel tempering} \cite{STrev}.)
In this method, a number of
non-interacting copies of the original system (or replicas)
at different temperatures are
simulated independently and
simultaneously by the conventional MC or MD method. Every few steps,
pairs of replicas are exchanged with a specified transition
probability.
The weight factor is just the
product of Boltzmann factors, and so it is essentially known.

REM has already been used in many applications in protein 
systems \cite{H97,FD,IRB2,SO,Kol,SKO,Gar}. 
Systems of Lennard-Jones particles have also been studied by
this method in various ensembles \cite{YP}--\cite{OKOM}.
Moreover, REM was applied to cluster studies in quantum
chemistry field \cite{ISNO}.
The details of molecular dynamics algorithm have been worked out
for REM \cite{SO} (see also Refs.~\cite{H97,Yama}).
We then developed a multidimensional REM which is particularly
useful in free energy calculations \cite{SKO}
(see also Refs.~\cite{Huk2,YP,Dunw}).

However, REM also has a computational difficulty:
As the number of degrees of freedom of the system increases,
the required number of replicas also greatly increases, whereas 
only a single replica is simulated in MUCA or ST.
This demands a lot of computer power for complex systems.
Our solution to this problem is: Use REM for the weight
factor determinations of MUCA or ST, which is much
simpler than previous iterative methods of weight
determinations, and then perform a long MUCA or ST
production run.
The first example is
the {\it replica-exchange multicanonical algorithm} (REMUCA)
\cite{SO3}.
In REMUCA,
a short replica-exchange simulation is performed, and the multicanonical
weight factor is determined by
the multiple-histogram reweighting techniques \cite{FS2,WHAM}.
Another example of such a combination is the
{\it replica-exchange simulated tempering} (REST) \cite{MO4}.
In REST, a short replica-exchange simulation is performed, and
the simulated tempering weight factor is determined by
the multiple-histogram reweighting techniques \cite{FS2,WHAM}.

We have introduced a further extension of REMUCA,
 which we refer to as {\it multicanonical replica-exchange method}
 (MUCAREM) \cite{SO3}.
In MUCAREM, the multicanonical weight factor is first
determined as in REMUCA, and then
 a replica-exchange multicanonical production simulation is performed
 with a small number of replicas.

In this article, we describe the six generalized-ensemble algorithms
mentioned above.  Namely, we first review three familiar methods:
MUCA, ST, and REM.  We then present the three new algorithms:
REMUCA, REST, and MUCAREM.
The effectiveness of these methods is tested with short
peptide systems.

\section{GENERALIZED-ENSEMBLE ALGORITHMS}

\subsection{Multicanonical Algorithm and Simulated Tempering}

Let us consider a system of $N$ atoms of 
mass $m_k$ ($k=1, \cdots, N$)
with their coordinate vectors and
momentum vectors denoted by 
$q \equiv \{{\biq}_1, \cdots, {\biq}_N\}$ and 
$p \equiv \{{\bip}_1, \cdots, {\bip}_N\}$,
respectively.
The Hamiltonian $H(q,p)$ of the system is the sum of the
kinetic energy $K(p)$ and the potential energy $E(q)$:
\begin{equation}
H(q,p) =  K(p) + E(q)~,
\label{eqn1}
\end{equation}
where
\begin{equation}
K(p) =  \sum_{k=1}^N \frac{{\bip_k}^2}{2 m_k}~.
\label{eqn2}
\end{equation}

In the canonical ensemble at temperature $T$ 
each state $x \equiv (q,p)$ with the Hamiltonian $H(q,p)$
is weighted by the Boltzmann factor:
\begin{equation}
W_{\rm B}(x;T) = e^{-\beta H(q,p)}~,
\label{eqn3}
\end{equation}
where the inverse temperature $\beta$ is defined by 
$\beta = 1/k_{\rm B} T$ ($k_{\rm B}$ is the Boltzmann constant). 
The average kinetic energy at temperature $T$ is then given by
\begin{equation}
\left< ~K(p)~ \right>_T =  
\left< \sum_{k=1}^N \frac{{\bip_k}^2}{2 m_k} \right>_T 
= \frac{3}{2} N k_{\rm B} T~.
\label{eqn4}
\end{equation}

Because the coordinates $q$ and momenta $p$ are decoupled
in Eq.~(\ref{eqn1}), we can suppress the kinetic energy
part and can write the Boltzmann factor as
\begin{equation}
W_{\rm B}(x;T) = W_{\rm B}(E;T) = e^{-\beta E}~.
\label{eqn4b}
\end{equation}
The canonical probability distribution of potential energy
$P_{\rm B}(E;T)$ is then given by the product of the
density of states $n(E)$ and the Boltzmann weight factor
$W_{\rm B}(E;T)$:
\begin{equation}
 P_{\rm B}(E;T) \propto n(E) W_{\rm B}(E;T)~.
\label{eqn4c}
\end{equation}
Since $n(E)$ is a rapidly
increasing function and the Boltzmann factor decreases exponentially,
the canonical ensemble yields a bell-shaped distribution
which has a maximum around the average energy at temperature $T$. The
conventional MC or MD simulations at constant temperature
are expected to yield $P_{\rm B}(E;T)$,
but, in practice, it is very difficult to obtain accurate canonical
distributions of complex systems at low temperatures by
conventional simulation methods.
This is because simulations at low temperatures tend to get
trapped in one or a few of local-minimum-energy states.

In the multicanonical ensemble (MUCA) \cite{MUCA1,MUCA2}, on the other
hand, each state is weighted by
a non-Boltzmann weight
factor $W_{\rm mu}(E)$ (which we refer to as the {\it multicanonical
weight factor}) so that a uniform energy
distribution $P_{\rm mu}(E)$ is obtained:
\begin{equation}
 P_{\rm mu}(E) \propto n(E) W_{\rm mu}(E) \equiv {\rm constant} .
\label{eqn5}
\end{equation}
The flat distribution implies that
a free random walk in the potential energy space is realized
in this ensemble.
This allows the simulation to escape from any local minimum-energy states
and to sample the configurational space much more widely than the conventional
canonical MC or MD methods.

From the definition in Eq.~(\ref{eqn5}) the multicanonical
weight factor is inversely proportional to the density of
states, and we can write it as follows:
\begin{equation}
 W_{\rm mu}(E) \equiv e^{-\beta_0 E_{\rm mu}(E;T_0)}
= \frac{1}{n(E)}~,
\label{eqn6}
\end{equation}
where we have chosen an arbitrary reference
temperature, $T_0 = 1/k_{\rm B} \beta_0$, and
the ``{\it multicanonical potential energy}''
is defined by
\begin{equation}
 E_{\rm mu}(E;T_0) = k_{\rm B} T_0 \ln n(E) = T_0 S(E)~.
\label{eqn7}
\end{equation}
Here, $S(E)$ is the entropy in the microcanonical
ensemble.
Since the density of states of the system is usually unknown,
the multicanonical weight factor
has to be determined numerically by iterations of short preliminary
runs \cite{MUCA1,MUCA2} as described in detail below.

A multicanonical Monte Carlo simulation
is performed, for instance, with the usual Metropolis criterion \cite{Metro}:
The transition probability of state $x$ with potential energy
$E$ to state $x^{\prime}$ with potential energy $E^{\prime}$ is given by
\begin{equation}
w(x \rightarrow x^{\prime}) = \left\{
\begin{array}{ll}
 1~, & {\rm for} \ \Delta E_{\rm mu} \le 0~, \cr
 \exp \left( - \beta_0 \Delta E_{\rm mu} \right)~, & {\rm for} \
\Delta E_{\rm mu} > 0~,
\end{array}
\right.
\label{eqn8}
\end{equation}
where
\begin{equation}
\Delta E_{\rm mu} \equiv E_{\rm mu}(E^{\prime};T_0) - E_{\rm mu}(E;T_0)~.
\label{eqn9}
\end{equation}
The molecular dynamics algorithm in multicanonical ensemble
also naturally follows from Eq.~(\ref{eqn6}), in which the
regular constant temperature molecular dynamics simulation
(with $T=T_0$) is performed by solving the following modified
Newton equation: \cite{HOE96,NNK} 
\begin{equation}
\dot{{\bip}}_k ~=~ - \frac{\partial E_{\rm mu}(E;T_0)}{\partial {\biq}_k}
~=~ \frac{\partial E_{\rm mu}(E;T_0)}{\partial E}~{\bif}_k~,
\label{eqn9a}
\end{equation}
where ${\bif}_k$ is the usual force acting on the $k$-th atom
($k = 1, \cdots, N$).
From Eq.~(\ref{eqn7}) this equation can be rewritten as
\begin{equation}
\dot{{\bip}}_k ~=~ \frac{T_0}{T(E)}~{\bif}_k~,
\label{eqn9b}
\end{equation}
where the following thermodynamic relation gives
the definition of the ``effective temperature''
$T(E)$:
\begin{equation}
\left. \frac{\partial S(E)}{\partial E}\right|_{E=E_a}
~=~ \frac{1}{T(E_a)}~,
\label{eqn9c}
\end{equation}
with
\begin{equation}
E_a ~=~ <E>_{T(E_a)}~.
\label{eqn9d}
\end{equation}

The multicanonical weight factor is usually determined by
iterations of short trial simulations.  The details of this process
are described, for instance, in Refs.~\cite{MUCA3,OH}.
For the first run, a canonical simulation at a sufficiently
high temperature $T_0$ is performed, i.e., we set
\begin{equation}
\left\{
\begin{array}{rl}
E^{(1)}_{\rm mu}(E;T_0) &=~ E~, \\
W^{(1)}_{\rm mu}(E;T_0) &=~ W_{\rm B} (E;T_0) ~=~ \exp \left(-\beta_0 E\right)~.
\end{array}
\right.
\label{iter}
\end{equation}
We define the maximum energy value $E_{\rm max}$ under which
we want to have a flat energy distribution by
the average potential energy at temperature $T_0$:
\begin{equation}
E_{\rm max} = <E>_{T_0}~.
\label{iter2}
\end{equation}
Above $E_{\rm max}$ we have the canonical distribution
at $T=T_0$.  In the $\ell$-th iteration a simulation with
the weight $W^{(\ell)}_{\rm mu}(E;T_0)=
\exp \left(-\beta_0 E^{(\ell)}_{\rm mu}(E;T_0)\right)$
is performed, and the histogram $N^{(\ell)}(E)$
of the potential energy distribution
$P^{(\ell)}_{\rm mu}(E)$ is obtained.
Let $E^{(\ell)}_{\rm min}$ be the lowest-energy value
that was obtained throughout the preceding iterations including
the present simulation.
The multicanonical weight factor for the $(\ell+1)$-th iteration is
then given by
\begin{equation}
E^{(\ell+1)}_{\rm mu} (E;T_0) = 
\left\{
   \begin{array}{@{\,}ll}
E~, & \mbox{for $E \ge E_{\rm max}$,} \\
E^{(\ell)}_{\rm mu} (E;T_0)
+ k_{\rm B} T_0 \ln N^{(\ell)}(E) - c^{(\ell)}~,
    & \mbox{for $E^{(\ell)}_{\rm min} \le E < E_{\rm max}$,} \\
   \left. \dis{\frac{\partial E^{(\ell+1)}_{\rm mu}(E;T_0)}{\partial E}}
        \right|_{E=E^{(\ell)}_{\rm min}} \left(E - E^{(\ell)}_{\rm min}\right)
             + E^{(\ell+1)}_{\rm mu}(E^{(\ell)}_{\rm min};T_0) , &
         \mbox{for $E < E^{(\ell)}_{\rm min}$,} 
\end{array}
\right.
\label{iter3}
\end{equation}
where the constant $c^{(\ell)}$ is introduced to ensure
the continuity at $E=E_{\rm max}$ and we have
\begin{equation}
c^{(\ell)} = k_{\rm B} T_0 \ln N^{(\ell)}(E_{\rm max})~. 
\label{iter4}
\end{equation}
We iterate this process until the obtained energy distribution
becomes reasonably flat, say, of the same order of magnitude,
for $E < E_{\rm max}$.  When the convergence is reached,
we should have that $E^{(\ell)}_{\rm min}$ is equal to
the global-minimum potential energy value. 

It is also common especially when working in MD algorithm to
use polynomials and other smooth functions to fit the histograms
during the iterations \cite{KI95,NNK,BK}.  We have shown that
the cubic spline functions work well \cite{SO3}.
 
However, the iterative process can be non-trivial and very tedius for
complex systems, and there have been attempts to accelerate
the convergence of the iterative process
\cite{MUCA3,KPV,SmBr,H97c,MUCAW,BK}.

After the optimal multicanonical weight factor is determined, one
performs a long multicanonical simulation once.  By monitoring
the potential energy throughout the simulation, one can find the
global-minimum-energy state.
Moreover, by using the obtained histogram $N_{\rm mu}(E)$ of the
potential energy distribution $P_{\rm mu}(E)$
the expectation value of a physical quantity $A$
at any temperature $T=1/k_{\rm B} \beta$
is calculated from
\begin{equation}
<A>_{T} \ = \frac{\dis{\sum_{E}~A(E)~n(E)~e^{-\beta E}}}
{\dis{\sum_{E} ~n(E)~e^{-\beta E}}}~,
\label{eqn18}
\end{equation}
where the best estimate of the
density of states is given by the single-histogram
reweighting techniques (see Eq.~(\ref{eqn5})) \cite{FS1}:
\begin{equation}
 n(E) = \frac{N_{\rm mu}(E)}{W_{\rm mu}(E)}~.
\label{eqn17}
\end{equation}
In the numerical work, we want to avoid round-off errors
(and overflows and underflows) as much as possible.
It is usually better to combine exponentials as follows
(see Eq.~(\ref{eqn6})):
\begin{equation}
<A>_{T} \ = 
\frac{\dis{\sum_{E}
~A(E)~N_{\rm mu}(E)~e^{\beta_0 E_{\rm mu}(E;T_0)-\beta E}}}
{\dis{\sum_{E} 
~N_{\rm mu}(E)~e^{\beta_0 E_{\rm mu}(E;T_0)-\beta E}}}~.
\label{eqn18b}
\end{equation}

We now briefly review the original {\it simulated tempering} (ST)
method \cite{ST1,ST2}.  In this method temperature itself becomes a
dynamical variable, and both the configuration and the temperature are updated
during the simulation with a weight:
\begin{equation}
W_{\rm ST} (E;T) = e^{-\beta E + a(T)}~,
\label{Eqn1}
\end{equation}
where the function $a(T)$ is chosen so that the probability distribution
of temperature is flat:
\begin{equation}
P_{\rm ST}(T) = \int dE~ n(E)~ W_{{\rm ST}} (E;T) =
\int dE~ n(E)~ e^{-\beta E + a(T)} = {\rm constant}~.
\lab{Eqn2}
\end{equation}
Hence, in simulated tempering the {\it temperature} is sampled
uniformly. A free random walk in temperature space
is realized, which in turn
induces a random walk in potential energy space and
allows the simulation to escape from
states of energy local minima.

In the numerical work we discretize the temperature in
$M$ different values, $T_m$ ($m=1, \cdots, M$).  Without loss of
generality we can order the temperature
so that $T_1 < T_2 < \cdots < T_M$.  The lowest temperature
$T_1$ should be sufficiently low so that the simulation can explore the
global-minimum-energy region, and
the highest temperature $T_M$ should be sufficiently high so that
no trapping in an energy-local-minimum state occurs.  The probability
weight factor in Eq.~(\ref{Eqn1}) is now written as
\begin{equation}
W_{\rm ST}(E;T_m) = e^{-\beta_m E + a_m}~,
\label{Eqn3}
\end{equation}
where $a_m=a(T_m)$ ($m=1, \cdots, M$).
The parameters $a_m$ are not known {\it a priori} and have to be determined
by iterations of short simulations.
This process can be non-trivial and very difficult for complex systems.
Note that from Eqs.~(\ref{Eqn2}) and (\ref{Eqn3}) we have
\begin{equation}
e^{-a_m} \propto \int dE~ n(E)~ e^{- \beta_m E}~.
\label{Eqn4}
\end{equation}
The parameters $a_m$ are therefore ``dimensionless'' Helmholtz free energy
at temperature $T_m$
(i.e., the inverse temperature $\beta_m$ multiplied by
the Helmholtz free energy).

Once the parameters $a_m$ are determined and the initial configuration and
the initial temperature $T_m$ are chosen,
a simulated tempering simulation is then realized by alternately
performing the following two steps \cite{ST1,ST2}:
\begin{enumerate}
\item A canonical MC or MD simulation at the fixed temperature $T_m$
is carried out for a certain MC or MD steps.
\item The temperature $T_m$ is updated to the neighboring values
$T_{m \pm 1}$ with the configuration fixed.  The transition probability of
this temperature-updating
process is given by the Metropolis criterion (see Eq.~(\ref{Eqn3})):
\begin{equation}
w(T_m \rightarrow T_{m \pm 1})
= \left\{
\begin{array}{ll}
 1~, & {\rm for} \ \Delta \le 0~, \cr
 \exp \left( - \Delta \right)~, & {\rm for} \ \Delta > 0~,
\end{array}
\right.
\label{Eqn5}
\end{equation}
where
\begin{equation}
\Delta = \left(\beta_{m \pm 1} - \beta_m \right) E
- \left(a_{m \pm 1} - a_m \right)~.
\label{Eqn6}
\end{equation}
\end{enumerate}
Note that in Step 2 we exchange only pairs of 
neighboring temperatures in order to secure sufficiently
large acceptance ratio of temperature updates.

As in multicanonical algorithm, the simulated tempering
parameters $a_m=a(T_m)$ ($m=1, \cdots, M$)
are also determined by iterations of short trial simulations
(see, e.g.,  Refs.~\cite{STrev,IRB1,HO96b} for details).
Here, we give the one in Ref.~\cite{HO96b}.

During the trial simulations we keep track of the temperature
distribution as a histogram $N_m=N(T_m)$ ($m=1, \cdots, M$).
\begin{enumerate}
\item Start with a short canonical simulation (i.e., $a_m=0$)
      updating only configurations at
      temperature $T_m = T_M$ (we initially set
      the temperature label $m$ to $M$)
      and calculate the average
      potential energy $<E>_{T_M}$.  
      Here, the histogram $N_n$ will have non-zero entry only
      for $n=m=M$.
\item Calculate new parameters $a_n$ according to
      \begin{equation}
      a_n =  \left\{ \begin{array}{ll}
             a_n - \ln N_n~, &  {\rm for}~~ m \le n \le M~, \cr
             a_n -
<E>_{T_m}\left(\beta_{m-1} - \beta_m\right)~, &
              {\rm for}~~ n = m - 1~, \cr
            - \infty~, & {\rm for}~~n < m - 1~.
                      \end{array} \right.
      \end{equation}
      This weight implies that the temperature will range between
      $T_{m-1}$ and $T_M$.
\item Start a new simulation, now updating both configurations and
temperatures, with weight $W_{\rm ST} (E;T_n) = e^{-\beta_n E + a_n}$
      and sample the
      distribution of temperatures $T_n$ in the histogram $N_n= N(T_n)$. 
      For $T = T_{m-1}$ calculate the average potential
      energy $<E>_{T_{m-1}}$.
\item If the histogram $N_n$ is approximately flat in the temperature range
      $T_{m-1} \le T_n \le T_M$, set $m = m - 1$.  Otherwise, leave
      $m$ unchanged.
\item Iterate the last three steps until the obtained temperature
      distribution $N_n$ becomes flat over the whole temperature range
      $\left[ T_1,T_M \right]$.
\end{enumerate}

After the optimal simulated tempering weight factor is determined,
one performs a long simulated tempering run once.
From the results of this production run, one can obtain
the canonical ensemble average of a physical quantity $A$ as a function 
of temperature from Eq.~(\ref{eqn18}), where the 
density of states is given by
the multiple-histogram reweighting techniques \cite{FS2,WHAM}
as follows.
Let $N_m(E)$ and $n_m$ be respectively
the potential-energy histogram and the total number of
samples obtained at temperature $T_m=1/k_{\rm B} \beta_m$
($m=1, \cdots, M$). 
The best estimate of the density of states is then given by \cite{FS2,WHAM}
\begin{equation}
n(E) = \frac{\dis{\sum_{m=1}^M ~g_m^{-1}~N_m(E)}}
{\dis{\sum_{m=1}^M ~g_m^{-1}~n_m~e^{f_m-\beta_m E}}}~,
\label{Eqn8a}
\end{equation}
where
\begin{equation}
e^{-f_m} = \sum_{E} ~n(E)~e^{-\beta_m E}~.
\label{Eqn8b}
\end{equation}
Here, $g_m = 1 + 2 \tau_m$,
and $\tau_m$ is the integrated
autocorrelation time at temperature $T_m$.
Note that
Eqs.~(\ref{Eqn8a}) and
(\ref{Eqn8b}) are solved self-consistently
by iteration \cite{FS2,WHAM} to obtain
the dimensionless Helmholtz free energy $f_m$
(and the density of states $n(E)$).
We remark that in the numeraical work, it is often more stable
to use the following equations instead of
Eqs.~(\ref{Eqn8a}) and
(\ref{Eqn8b}):
\begin{equation}
P_{\rm B}(E;T) = n(E) e^{-\beta E} = 
\frac{\dis{\sum_{m=1}^M ~g_m^{-1}~N_m(E)}}
{\dis{\sum_{m=1}^M ~g_m^{-1}~n_m~e^{f_m - (\beta_m-\beta) E}}}~,
\label{Eqn8A}
\end{equation}
where
\begin{equation}
e^{-f_m} = \sum_{E} ~P_{\rm B}(E;T_m)~.
\label{Eqn8B}
\end{equation}
The equations are solved iteratively as follows.
We can set all the $f_m$ ($m=1, \cdots, M$) to, e.g., zero initially.
We then use Eq.~(\ref{Eqn8A}) to obtain 
$P_{\rm B}(E;T_m)$ ($m=1, \cdots, M$), which are substituted into
Eq.~(\ref{Eqn8B}) to obtain next values of $f_m$, and so on.

\subsection{Replica-Exchange Method}

The {\it replica-exchange method} (REM) \cite{RE1}--\cite{RE3}
was developed as an extension of
simulated tempering \cite{RE1} (thus it is also referred to as
{\it parallel tempering} \cite{STrev})
(see, e.g.,  Ref.~\cite{SO} for a detailed
description of the algorithm).
The system for REM consists of 
$M$ {\it non-interacting} copies (or, replicas) 
of the original system in the canonical ensemble
at $M$ different temperatures $T_m$ ($m=1, \cdots, M$).
We arrange the replicas so that there is always
exactly one replica at each temperature.
Then there is  a one-to-one correspondence between replicas
and temperatures; the label $i$ ($i=1, \cdots, M$) for replicas 
is a permutation of 
the label $m$ ($m=1, \cdots, M$) for temperatures,
and vice versa:
\begin{equation}
\left\{
\begin{array}{rl}
i &=~ i(m) ~\equiv~ f(m)~, \cr
m &=~ m(i) ~\equiv~ f^{-1}(i)~,
\end{array}
\right.
\label{eq4b}
\end{equation}
where $f(m)$ is a permutation function of $m$ and
$f^{-1}(i)$ is its inverse.

Let $X = \left\{x_1^{[i(1)]}, \cdots, x_M^{[i(M)]}\right\} 
= \left\{x_{m(1)}^{[1]}, \cdots, x_{m(M)}^{[M]}\right\}$ 
stand for a ``state'' in this generalized ensemble.
The state $X$ is specified by the $M$ sets of 
coordinates $q^{[i]}$ and momenta $p^{[i]}$
of $N$ atoms in replica $i$ at temperature $T_m$:
\begin{equation}
x_m^{[i]} \equiv \left(q^{[i]},p^{[i]}\right)_m~.
\label{eq5}
\end{equation}

Because the replicas are non-interacting, the weight factor for
the state $X$ in
this generalized ensemble is given by
the product of Boltzmann factors for each replica (or at each
temperature):
\begin{equation}
W_{\rm REM}(X) = \exp \left\{- \dis{\sum_{i=1}^M \beta_{m(i)} 
H\left(q^{[i]},p^{[i]}\right) } \right\}
 = \exp \left\{- \dis{\sum_{m=1}^M \beta_m 
H\left(q^{[i(m)]},p^{[i(m)]}\right) }
 \right\}~,
\label{eq7}
\end{equation}
where $i(m)$ and $m(i)$ are the permutation functions in 
Eq.~(\ref{eq4b}).

We now consider exchanging a pair of replicas in the generalized
ensemble.  Suppose we exchange replicas $i$ and $j$ which are
at temperatures $T_m$ and $T_n$, respectively:  
\begin{equation}
X = \left\{\cdots, x_m^{[i]}, \cdots, x_n^{[j]}, \cdots \right\} 
\longrightarrow \ 
X^{\prime} = \left\{\cdots, x_m^{[j] \prime}, \cdots, x_n^{[i] \prime}, 
\cdots \right\}~. 
\label{eq8}
\end{equation}
Here, $i$, $j$, $m$, and $n$ are related by the permutation
functions in Eq.~(\ref{eq4b}),
and the exchange of replicas introduces a new 
permutation function $f^{\prime}$:
\begin{equation}
\left\{
\begin{array}{rl}
i &= f(m) \longrightarrow j=f^{\prime}(m)~, \cr
j &= f(n) \longrightarrow i=f^{\prime}(n)~. \cr
\end{array}
\right.
\label{eq8c}
\end{equation}

The exchange of replicas can be written in more detail as
\begin{equation}
\left\{
\begin{array}{rl}
x_m^{[i]} \equiv \left(q^{[i]},p^{[i]}\right)_m & \longrightarrow \ 
x_m^{[j] \prime} \equiv \left(q^{[j]},p^{[j] \prime}\right)_m~, \cr
x_n^{[j]} \equiv \left(q^{[j]},p^{[j]}\right)_n & \longrightarrow \ 
x_n^{[i] \prime} \equiv \left(q^{[i]},p^{[i] \prime}\right)_n~,
\end{array}
\right.
\label{eq9}
\end{equation}
where the definitions for $p^{[i] \prime}$ and $p^{[j] \prime}$
will be given below.
We remark that this process is equivalent to exchanging
a pair of temperatures $T_m$ and $T_n$ for the
corresponding replicas $i$ and $j$ as follows:
\begin{equation}
\left\{
\begin{array}{rl}
x_m^{[i]} \equiv \left(q^{[i]},p^{[i]}\right)_m & \longrightarrow \ 
x_n^{[i] \prime} \equiv \left(q^{[i]},p^{[i] \prime}\right)_n~, \cr
x_n^{[j]} \equiv \left(q^{[j]},p^{[j]}\right)_n & \longrightarrow \ 
x_m^{[j] \prime} \equiv \left(q^{[j]},p^{[j] \prime}\right)_m~.
\end{array}
\right.
\label{eq10}
\end{equation}

In the original implementation of the 
{\it replica-exchange method} (REM) \cite{RE1}--\cite{RE3},
Monte Carlo algorithm was used, and only the coordinates $q$
(and the potential energy
function $E(q)$)
had to be taken into account.  
In molecular dynamics algorithm, on the other hand, we also have to
deal with the momenta $p$.
We proposed the following momentum 
assignment in Eq.~(\ref{eq9}) (and in Eq.~(\ref{eq10})) \cite{SO}:
\begin{equation}
\left\{
\begin{array}{rl}
p^{[i] \prime} & \equiv \dis{\sqrt{\frac{T_n}{T_m}}} ~p^{[i]}~, \cr
p^{[j] \prime} & \equiv \dis{\sqrt{\frac{T_m}{T_n}}} ~p^{[j]}~,
\end{array}
\right.
\label{eq11}
\end{equation}
which we believe is the simplest and the most natural.
This assignment means that we just rescale uniformly 
the velocities of all the atoms 
in the replicas by
the square root of the ratio of the two temperatures so that
the temperature condition in Eq.~(\ref{eqn4}) may be satisfied.

In order for this exchange process to converge towards an equilibrium
distribution, it is sufficient to impose the detailed balance
condition on the transition probability $w(X \rightarrow X^{\prime})$:
\begin{equation}
W_{\rm REM}(X) \  w(X \rightarrow X^{\prime})
= W_{\rm REM}(X^{\prime}) \  w(X^{\prime} \rightarrow X)~.
\label{eq12}
\end{equation}
From Eqs.~(\ref{eqn1}), (\ref{eqn2}), (\ref{eq7}), (\ref{eq11}), 
and (\ref{eq12}), we have
\begin{equation}
\begin{array}{rl}
\dis{\frac{w(X \rightarrow X^{\prime})} 
     {w(X^{\prime} \rightarrow X)}} 
&= \exp \left\{ 
- \beta_m \left[K\left(p^{[j] \prime}\right) + E\left(q^{[j]}\right)\right] 
- \beta_n \left[K\left(p^{[i] \prime}\right) + E\left(q^{[i]}\right)\right]
\right. \cr
& \ \ \ \ \ \ \ \ \ \  \left.
+ \beta_m \left[K\left(p^{[i]}\right) + E\left(q^{[i]}\right)\right] 
+ \beta_n \left[K\left(p^{[j]}\right) + E\left(q^{[j]}\right)\right]
\right\}~, \cr
&= \exp \left\{ 
- \beta_m \dis{\frac{T_m}{T_n}} K\left(p^{[j]}\right)
- \beta_n \dis{\frac{T_n}{T_m}} K\left(p^{[i]}\right)
+ \beta_m K\left(p^{[i]}\right)
+ \beta_n K\left(p^{[j]}\right)
\right. \cr
& \ \ \ \ \ \ \ \ \ \  \left.
- \beta_m \left[E\left(q^{[j]}\right)
                - E\left(q^{[i]}\right)\right] 
- \beta_n \left[E\left(q^{[i]}\right)
                - E\left(q^{[j]}\right)\right] 
\right\}~, \cr
&= \exp \left( - \Delta \right)~,
\end{array}
\label{eq13}
\end{equation}
where
\begin{equation}
\Delta \equiv \left(\beta_n - \beta_m \right)
              \left(E\left(q^{[i]}\right)
                  - E\left(q^{[j]}\right)\right)~, 
\label{eq14}
\end{equation}
and $i$, $j$, $m$, and $n$ are related by the permutation
functions (in Eq.~(\ref{eq4b})) before the exchange:
\begin{equation}
\left\{
\begin{array}{ll}
i &= f(m)~, \cr
j &= f(n)~.
\end{array}
\right.
\label{eq13b}
\end{equation}
This can be satisfied, for instance, by the usual Metropolis criterion
\cite{Metro}:
\begin{equation}
w(X \rightarrow X^{\prime}) \equiv
w\left( x_m^{[i]} ~\left|~ x_n^{[j]} \right. \right) 
= \left\{
\begin{array}{ll}
 1~, & {\rm for} \ \Delta \le 0~, \cr
 \exp \left( - \Delta \right)~, & {\rm for} \ \Delta > 0~,
\end{array}
\right.
\label{eq15}
\end{equation}
where in the second expression 
(i.e., $w( x_m^{[i]} | x_n^{[j]} )$) 
we explicitly wrote the
pair of replicas (and temperatures) to be exchanged.
Note that this is exactly the same criterion that was originally
derived for Monte Carlo algorithm \cite{RE1}--\cite{RE3}.

Without loss of generality we can
again assume $T_1 < T_2 < \cdots < T_M$.
A simulation of the 
{\it replica-exchange method} (REM) \cite{RE1}--\cite{RE3}
is then realized by alternately performing the following two
steps:
\begin{enumerate}
\item Each replica in canonical ensemble of the fixed temperature 
is simulated $simultaneously$ and $independently$
for a certain MC or MD steps. 
\item A pair of replicas at neighboring temperatures,
say $x_m^{[i]}$ and $x_{m+1}^{[j]}$, are exchanged
with the probability
$w\left( x_m^{[i]} ~\left|~ x_{m+1}^{[j]} \right. \right)$ 
in Eq.~(\ref{eq15}).
\end{enumerate}
Note that in Step 2 we exchange only pairs of replicas corresponding to
neighboring temperatures, because
the acceptance ratio of the exchange decreases exponentially
with the difference of the two $\beta$'s (see Eqs.~(\ref{eq14})
and (\ref{eq15})).
Note also that whenever a replica exchange is accepted
in Step 2, the permutation functions in Eq.~(\ref{eq4b})
are updated.

The REM simulation is particularly suitable for parallel
computers.  Because one can minimize the amount of information
exchanged among nodes, it is best to assign each replica to
each node (exchanging pairs of temperature values among nodes
is much faster than exchanging coordinates and momenta).
This means that we keep track of the permutation function
$m(i;t)=f^{-1}(i;t)$ in Eq.~(\ref{eq4b}) as a function
of MC or MD step $t$ during the simulation.
After parallel canonical MC or MD simulations for a certain
steps (Step 1), $M/2$ pairs of
replicas corresponding to neighboring temperatures
are simulateneously exchanged (Step 2), and the pairing is alternated 
between the two possible choices, i.e., ($T_1,T_2$), ($T_3,T_4$), $\cdots$
and ($T_2,T_3$), ($T_4,T_5$), $\cdots$.

The major advantage of REM over other generalized-ensemble
methods such as multicanonical algorithm \cite{MUCA1,MUCA2}
and simulated tempering \cite{ST1,ST2}
lies in the fact that the weight factor 
is {\it a priori} known (see Eq.~(\ref{eq7})), while
in the latter algorithms the determination of the
weight factors can be very tedius and time-consuming.
A random walk in ``temperature space'' is
realized for each replica, which in turn induces a random
walk in potential energy space.  This alleviates the problem
of getting trapped in states of energy local minima.
In REM, however, the number of required replicas increases
as the system size $N$ increases (according to $\sqrt N$) \cite{RE1}.
This demands a lot of computer power for complex systems.

\subsection{Replica-Exchange Multicanonical Algorithm and
Replica-Exchange Simulated Tempering} 

The {\it replica-exchange multicanonical algorithm} (REMUCA) 
\cite{SO3} overcomes
both the difficulties of MUCA (the multicanonical weight factor
determination is non-trivial)
and REM (a lot of replicas, or computation time, is required).
In REMUCA we first perform a short REM simulation (with $M$ replicas)
to determine the
multicanonical weight factor and then perform with this weight
factor a regular multicanonical simulation with high statistics.
The first step is accomplished by the multiple-histogram reweighting
techniques \cite{FS2,WHAM}.
Let $N_m(E)$ and $n_m$ be respectively
the potential-energy histogram and the total number of
samples obtained at temperature $T_m=1/k_{\rm B} \beta_m$ of the REM run.
The density of states $n(E)$ is then given by solving 
Eqs.~(\ref{Eqn8a}) and (\ref{Eqn8b}) self-consistently by iteration
\cite{FS2,WHAM}.

Once the estimate of the density of states is obtained, the
multicanonical weight factor can be directly determined from
Eq.~(\ref{eqn6}) (see also Eq.~(\ref{eqn7})).
Actually, the multicanonical potential energy, $E_{\rm mu}(E;T_0)$,
thus determined is only reliable in the following range:
\begin{equation}
E_1 \le E \le E_M~,
\label{eqn29}
\end{equation}
where 
\begin{equation}
\left\{
\begin{array}{rl}
E_1 &=~ <E>_{T_1}~, \\
E_M &=~ <E>_{T_M}~,
\end{array}
\right.
\label{eqn29b}
\end{equation}
and $T_1$ and $T_M$ are respectively the lowest and the highest
temperatures used in the REM run.
Outside this range we extrapolate
the multicanonical potential energy linearly:
\begin{equation}
 {\cal E}_{\rm mu}^{\{0\}}(E) \equiv \left\{
   \begin{array}{@{\,}ll}
   \left. \dis{\frac{\partial E_{\rm mu}(E;T_0)}{\partial E}}
        \right|_{E=E_1} (E - E_1)
             + E_{\rm mu}(E_1;T_0)~, &
         \mbox{for $E < E_1$,} \\
         E_{\rm mu}(E;T_0)~, &
         \mbox{for $E_1 \le E \le E_M$,} \\
   \left. \dis{\frac{\partial E_{\rm mu}(E;T_0)}{\partial E}}
        \right|_{E=E_M} (E - E_M)
             + E_{\rm mu}(E_M;T_0)~, &
         \mbox{for $E > E_M$.}
   \end{array}
   \right.
\label{eqn31}
\end{equation}
The multicanonical MC and MD runs are then performed with
the Metropolis criterion of Eq.~(\ref{eqn8})
and with
the Newton equation in Eq.~(\ref{eqn9a}), respectively,
in which 
${\cal E}_{\rm mu}^{\{0\}}(E)$ in
Eq.~(\ref{eqn31}) is substituted into $E_{\rm mu}(E;T_0)$.
We expect to obtain a flat potential energy distribution in
the range of Eq.~(\ref{eqn29}).
Finally, the results are analyzed by the single-histogram
reweighting techniques as described in Eq.~(\ref{eqn17})
(and Eq.~(\ref{eqn18})).

Some remarks are now in order.
From Eqs.~(\ref{eqn7}), (\ref{eqn9c}), (\ref{eqn9d}),
and (\ref{eqn29b}), Eq.~(\ref{eqn31}) becomes
\begin{equation}
 {\cal E}_{\rm mu}^{\{0\}}(E) = \left\{
   \begin{array}{@{\,}ll}
    \dis{\frac{T_0}{T_1}} (E - E_1) + T_0 S(E_1) = 
    \dis{\frac{T_0}{T_1}} E + {\rm constant}~, &
         \mbox{for $E < E_1 \equiv <E>_{T_1}$,} \\
         T_0 S(E)~, &
         \mbox{for $E_1 \le E \le E_M$,} \\
    \dis{\frac{T_0}{T_M}} (E - E_M) + T_0 S(E_M) = 
    \dis{\frac{T_0}{T_M}} E + {\rm constant}~, &
         \mbox{for $E > E_M \equiv <E>_{T_M}$.}
   \end{array}
   \right.
\label{eqn31b}
\end{equation}
The Newton equation in Eq.~(\ref{eqn9a}) is then written as
(see Eqs.~({eqn9b}), (\ref{eqn9c}), and (\ref{eqn9d}))
\begin{equation}
\dot{{\bip}}_k = \left\{
   \begin{array}{@{\,}ll}
   \dis{\frac{T_0}{T_1}}~{\bif}_k~, &
         \mbox{for $E < E_1$,} \\
   \dis{\frac{T_0}{T(E)}}~{\bif}_k~, &
         \mbox{for $E_1 \le E \le E_M$,} \\
   \dis{\frac{T_0}{T_M}}~{\bif}_k~, &
         \mbox{for $E > E_M$.}
   \end{array}
   \right.
\label{eqn31c}
\end{equation}
Because only the product of inverse temperature $\beta$ and
potential energy $E$ enters in the Boltzmann factor
(see Eq.~(\ref{eqn4b})), a rescaling of the potential energy
(or force) by a constant, say $\alpha$, can be considered as
the rescaling of the temperature by $\alpha^{-1}$ \cite{HOE96,Yama}.  
Hence,
our choice of ${\cal E}_{\rm mu}^{\{0\}}(E)$
in Eq.~(\ref{eqn31}) results in a canonical simulation at
$T=T_1$ for $E < E_1$, a multicanonical simulation for
$E_1 \le E \le E_M$, and a canonical simulation at
$T=T_M$ for $E > E_M$.
Note also that the above arguments are independent of
the value of $T_0$, and we
will get the same results, regardless of its value.

Finally, although we did not find any difficulty in the case 
of protein systems that we studied,
a single REM run in general may not be able to
give an accurate estimate of the
density of states (like in the case of
a first-order phase transition \cite{RE1}).  In such a
case we can still greatly simplify the process of the
multicanonical weight factor determination by
combining the present method with the
previous iterative methods \cite{MUCA3,OH,KPV,SmBr,H97c,MUCAW,BK}.

We finally present the new method which we refer to as the 
{\it replica-exchange simulated tempering} (REST) \cite{MO4}.  
In this method, just as in REMUCA,
we first perform a short REM simulation (with $M$ replicas)
to determine the simulated tempering
weight factor and then perform with this weight
factor a regular ST simulation with high statistics.
The first step is accomplished by 
the multiple-histogram reweighting
techniques \cite{FS2,WHAM}, which give
the dimensionless Helmholtz free energy $f_m$ (see Eqs.~(\ref{Eqn8a})
and (\ref{Eqn8b})).

Once the estimate of the dimensionless Helmholtz free energy $f_m$ are
obtained, the simulated tempering 
weight factor can be directly determined by using
Eq.~(\ref{Eqn3}) where we set $a_m = f_m$ (compare Eqs.~(\ref{Eqn4})
and (\ref{Eqn8b})).
A long simulated tempering run is then performed with this
weight factor.  
Let $N_m(E)$ and $n_m$ be respectively
the potential-energy histogram and the total number of
samples obtained at temperature $T_m=1/k_{\rm B} \beta_m$ from this
simulated tempering run.  The multiple-histogram
reweighting techniques of Eqs.~(\ref{Eqn8a}) and (\ref{Eqn8b}) can be used
again to obtain the best estimate of the density of states
$n(E)$.
The expectation value of a physical quantity $A$
at any temperature $T~(= 1/k_{\rm B} \beta)$ is then calculated from
Eq.~(\ref{eqn18}).

The formulations of REMUCA and REST are simple and straightforward, but
the numerical improvement is great, because the weight factor
determination for MUCA and ST becomes very difficult
by the usual iterative processes for complex systems.

\subsection{Multicanonical Replica-Exchange Method}

In the previous subsection we presented a new generalized-ensemble
algorithm, REMUCA, that combines the merits of replica-exchange method
and multicanonical algorithm.  In REMUCA a short REM simulation with
$M$ replicas are first performed and the results are used to determine
the multicanonical weight factor, and then a regular multicanonical
production run with this weight is performed.
The number of replicas, $M$, that is required in the first step should
be set minimally as long as a random walk between 
the lowest-energy region and the
high-energy region is realized.  This number can still be
very large for complex systems.  This is why the (multicanonical)
production run in REMUCA is performed with a ``single replica.'' 
While multicanonical simulatoins are usually based on local
updates, a replica-exchange process can be considered to be a
global update, and global updates enhance the sampling further.
Here, we present a further modification of REMUCA and refer to the
new method as {\it multicanonical replica-exchange method}
(MUCAREM) \cite{SO3}.  In MUCAREM the final production run is not a 
regular multicanonical simulation but a replica-exchange simulation
with a few replicas, say ${\cal M}$ replicas,
in the multicanonical ensemble. 
(We remark that replica-exchange simulations based on the
generalized ensemble with Tsallis weights
were introduced in Ref.~\cite{H97}.)
Because multicanonical simulations cover much wider energy
ranges than regular canonical simulations, the number of
required replicas for the production run of MUCAREM is
much less than that for the regular REM (${\cal M} \ll M$),
and we can keep the merits of
REMUCA (and improve the sampling further).

The details of MUCAREM are as follows.
As in REMUCA, we first perform a short REM simulation with $M$ replicas
with $M$ different temperatures (we order them as 
$T_1 < T_2 < \cdots < T_M$)
and obtain the best estimate of the density of states $n(E)$
in the whole energy range of interest (see Eq.~(\ref{eqn29}))
by the multiple-histogram
reweighting techniques of Eqs.~(\ref{Eqn8a}) and (\ref{Eqn8b}).
We then choose a number ${\cal M}$ (${\cal M} \ll M$) and
assign ${\cal M}$ pairs of temperatures 
($T_{\rm L}^{\{m\}}, T_{\rm H}^{\{m\}}$) ($m = 1, \cdots, {\cal M}$).
Here, we assume that
$T_{\rm L}^{\{m\}} < T_{\rm H}^{\{m\}}$ and arrange the temperatures 
so that the neighboring regions covered by the pairs
have sufficient overlaps.  In particular, we set  
$T_{\rm L}^{\{1\}} = T_1$ and $T_{\rm H}^{\{{\cal M}\}} = T_M$.
We then define the following quantities:
\begin{equation}
\left\{
\begin{array}{rl}
E_{\rm L}^{\{m\}} &=~ <E>_{T_{\rm L}^{\{m\}}}~, \\
E_{\rm H}^{\{m\}} &=~ <E>_{T_{\rm H}^{\{m\}}}~,~(m = 1, \cdots, {\cal M})~.
\end{array}
\right.
\label{eqn32}
\end{equation}
We also choose ${\cal M}$ (arbitrary) temperatures
$T_m$ ($m = 1, \cdots, {\cal M}$) and assign 
the following multicanonical potential energies:
\begin{equation}
 {\cal E}_{\rm mu}^{\{m\}}(E) = \left\{
   \begin{array}{@{\,}ll}
   \left. \dis{\frac{\partial E_{\rm mu}(E;T_m)}{\partial E}}
        \right|_{E=E_{\rm L}^{\{m\}}} (E - E_{\rm L}^{\{m\}})
             + E_{\rm mu}(E_{\rm L}^{\{m\}};T_m)~, &
         \mbox{for $E < E_{\rm L}^{\{m\}}$,} \\
         E_{\rm mu}(E;T_m)~, &
         \mbox{for $E_{\rm L}^{\{m\}} \le E \le E_{\rm H}^{\{m\}}$,} \\
   \left. \dis{\frac{\partial E_{\rm mu}(E;T_m)}{\partial E}}
        \right|_{E=E_{\rm H}^{\{m\}}} (E - E_{\rm H}^{\{m\}})
             + E_{\rm mu}(E_{\rm H}^{\{m\}};T_m)~, &
         \mbox{for $E > E_{\rm H}^{\{m\}}$,}
   \end{array}
   \right. 
\label{eqn33}
\end{equation}
where $E_{\rm mu}(E;T)$ is the multicanonical potential energy 
that was determined for the whole energy range
of Eq.~(\ref{eqn29}).
As remarked around Eq.~(\ref{eqn31b}),
our choice of ${\cal E}_{\rm mu}^{\{m\}}(E)$ 
in Eq.~(\ref{eqn33}) results in a canonical simulation at
$T=T_{\rm L}^{\{m\}}$ for $E < E_{\rm L}^{\{m\}}$, a multicanonical 
simulation for
$E_{\rm L}^{\{m\}} \le E \le E_{\rm H}^{\{m\}}$, and a canonical simulation at
$T=T_{\rm H}^{\{m\}}$ for $E > E_{\rm H}^{\{m\}}$.

The production run of MUCAREM is a replica-exchange simulation
with ${\cal M}$ replicas with ${\cal M}$ different
temperatures $T_m$ and multicanonical potential
energies ${\cal E}_{\rm mu}^{\{m\}}(E)$.
By following the same derivation that led to the original
REM, we have the following 
transition probability of replica exchange of
neighboring temperatures (see Eqs.~(\ref{eq14}) and (\ref{eq15})):
\begin{equation}
w\left( x_m^{[i]} ~\left|~ x_{m+1}^{[j]} \right. \right)
= \left\{
\begin{array}{ll}
 1~, & {\rm for} \ \Delta \le 0~, \cr
 \exp \left( - \Delta \right)~, & {\rm for} \ \Delta > 0~,
\end{array}
\right.
\label{Eqn20}
\end{equation}
where
\begin{equation}
\Delta = \beta_{m+1}
\left\{{\cal E}_{\rm mu}^{\{m+1\}}\left(E\left(q^{[i]}\right)\right) -
{\cal E}_{\rm mu}^{\{m+1\}}\left(E\left(q^{[j]}\right)\right)\right\}
- \beta_{m}
\left\{{\cal E}_{\rm mu}^{\{m\}}\left(E\left(q^{[i]}\right)\right) -
{\cal E}_{\rm mu}^{\{m\}}\left(E\left(q^{[j]}\right)\right)\right\}~.
\label{Eqn21}
\end{equation}
Note that we need to newly evaluate the multicanonical
potential energy, ${\cal E}_{\rm mu}^{\{m\}}(E(q^{[j]}))$ and
${\cal E}_{\rm mu}^{\{m+1\}}(E(q^{[i]}))$, because 
${\cal E}_{\rm mu}^{\{m\}}(E)$ and
${\cal E}_{\rm mu}^{\{n\}}(E)$ are, in
general, different functions for $m \ne n$. 
We remark that the same additional evaluation of the
potential energy is necessary for the multidimensional replica-exchange
method \cite{SKO}.

For obtaining the canonical distributions,
the multiple-histogram reweighting techniques \cite{FS2,WHAM}
are again used.
Let $N_m(E)$ and $n_m$ be respectively
the potential-energy histogram and the total number of
samples obtained at $T_m$ with the multicanonical
potential energy ${\cal E}_{\rm mu}^{\{m\}}(E)$
($m = 1, \cdots, {\cal M}$).
The expectation value
of a physical quantity $A$ 
at any temperature $T=1/k_{\rm B} \beta$
is then obtained from Eq.~(\ref{eqn18}),
where the best estimate of the density of states is given by
solving the multiple-histogram reweighting equations,
which now read
\begin{equation}
n(E) = \frac{\dis{\sum_{m=1}^{\cal M} ~g_m^{-1}~N_m(E)}}
{\dis{\sum_{m=1}^{\cal M} 
~g_m^{-1}~n_m~e^{f_m-\beta_m {\cal E}_{\rm mu}^{\{m\}}(E)}}}~,
\label{eqn41}
\end{equation}
and
\begin{equation}
e^{-f_m} = \sum_{E} ~n(E)~e^{-\beta_m {\cal E}_{\rm mu}^{\{m\}}(E)}~.
\label{eqn42}
\end{equation}

\section{EXAMPLES OF SIMULATION RESULTS}
   
We now present some examples of the simulation results by
the algorithms described in the previous section.
A few short peptide systems were considered.

For Monte Carlo simulations, 
the potential energy parameters were taken from
ECEPP/2 \cite{ECEP1}--\cite{ECEP3}.
The generalized-ensemble algorithms were implemented in
the computer code KONF90 \cite{KONF1,KONF2} for the
actual simulations.
Besides gas phase simulations, various solvation models
have been incorporated.  
The simplest one is the sigmoidal, distance-dependent
dielectric function \cite{SIG1,SIG2}.  The explicit form
of the function we used is given in Ref.~\cite{O2},
which is a slight modification of the one
in Ref.~\cite{DKK}.
A second (and more accurate) model that represents solvent contributions
is the term proportional to the solvent-accessible
surface area of solute molecule.
The parameters we used are those
of Ref.~\cite{OONS}.
For the calculation of solvent-accessible surface area, we used
the computer code
NSOL \cite{SOL2}, which is based on the code NSC \cite{SOL3}.
The third (and most rigorous) method that represents solvent
effects is based on the reference interaction
site model (RISM) \cite{RISM1}--\cite{RISM3}.
The model of water molecule that we adopted is the SPC/E
model \cite{SPC}.
A robust and fast algorithm for solving RISM
equations was recently developed \cite{KINO}, which
we employed in our calculations.

For molecular dynamics simulations,
the force-field parameters were taken from the all-atom
versions of AMBER \cite{AMBER1}--\cite{AMBER3}.
The computer code developed in
Refs. \cite{SK,KHG}, which is based on PRESTO
\cite{PRESTO}, was used. The unit time step was set to 0.5 fs.
The temperature during the canonical MD simulations was 
controlled by the constraint method \cite{HLM,EM}.
Besides gas phase simulations, we have also performed
MD simulations with
explicit water molecules of TIP3P model \cite{TIP3P}.

As described in detail in the previous section, in 
generalized-ensemble simulations and subsequent
analyses of the data, potential
energy distributions have to be taken as histograms.
For the bin size of these histograms, we used the values
ranging from 0.5 to 2 kcal/mol, depending on the system
studied.

We first illustrate how effectively generalized-ensemble
simulations can sample the configurational space compared
to the conventional simulations in the canonical ensemble.
It is known by experiments that the system of a 17-residue peptide fragment from 
ribonuclease T1 tends to form $\alpha$-helical
conformations \cite{Scholtz}.  We have performed both a
canonical MC simulation of this peptide at a low 
temperature ($T=200$ K) and a
multicanonical MC simulation \cite{MO5}.  In Figure 1
we show the time series of potential energy from these
simulations.

\begin{figure}[hbtp]
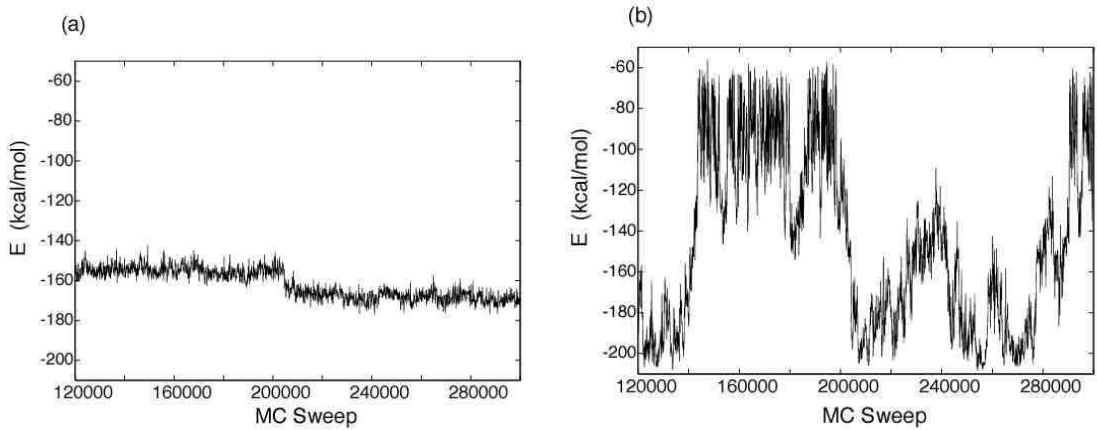

\begin{center}
\includegraphics[width=7.5cm,keepaspectratio]{bpfig1a.epsf}
\includegraphics[width=7.5cm,keepaspectratio]{bpfig1b.epsf}
\end{center}
\caption{Time series (from 120,000 MC sweeps to
300,000 MC sweeps) of potential energy of the
peptide fragment of ribonuclease T1 from (a)
a conventional canonical MC simulation at $T=200$ K
and (b) a multicanonical MC simulation.}
\label{fig1}
\end{figure}

We see that the canonical
simulation thermalize very slowly.
On the other hand, the MUCA simulation indeed performed
a random walk in potential energy space covering a
very wide energy range.  Four conformations
chosen during this period (from 120,000 MC sweeps to
300,000 MC sweeps) are shown in Figure 2 for the 
canonical simulation and
in Figure 3 for the MUCA simulation.
We see that the MUCA simulation samples much wider 
conformational space than the conventional canonical
simulation.

\begin{figure}[hbtp]
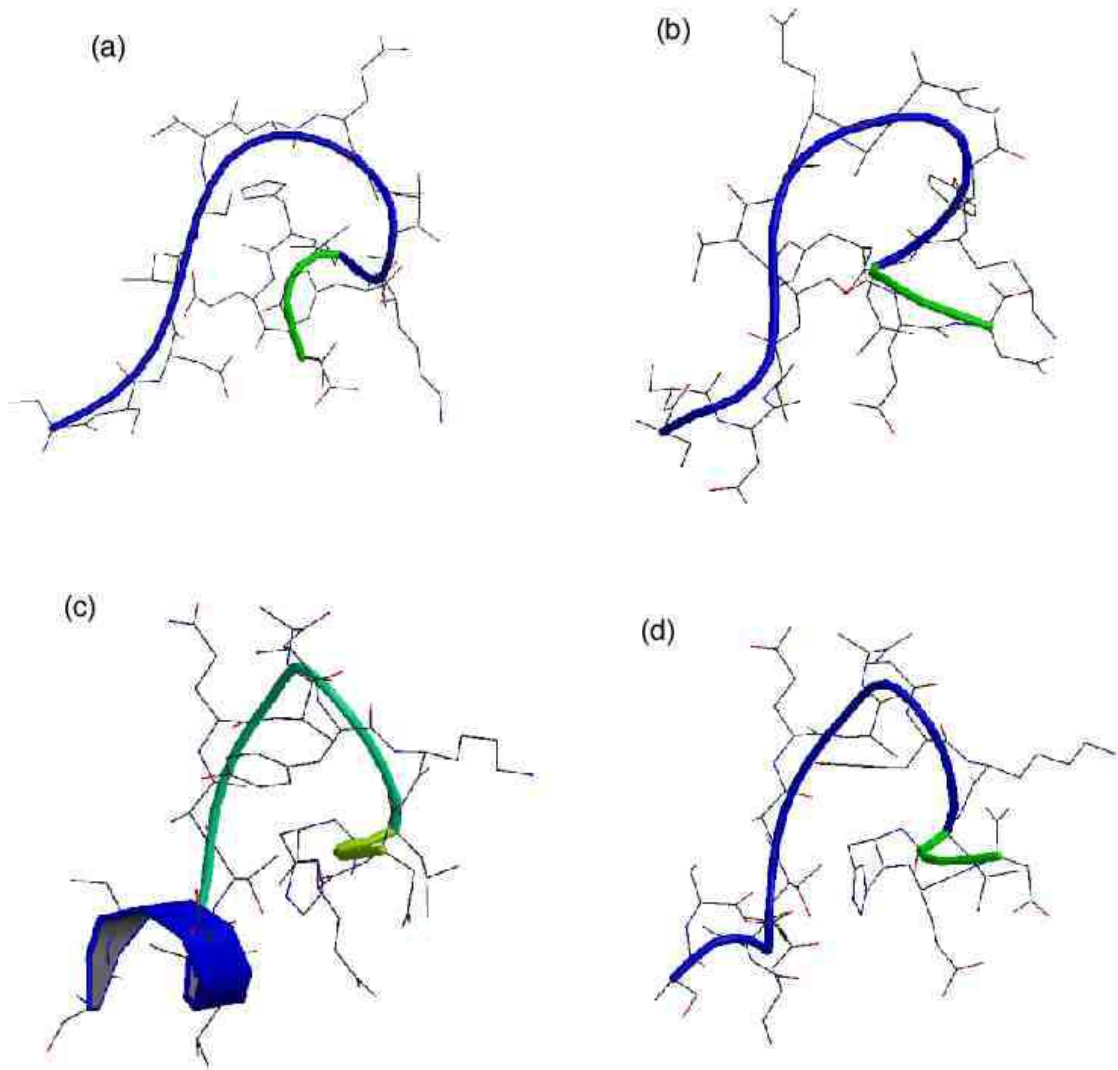

\begin{center}
\includegraphics[width=7.5cm,keepaspectratio]{bpfig2a.epsf}
\includegraphics[width=7.5cm,keepaspectratio]{bpfig2b.epsf}
\includegraphics[width=7.5cm,keepaspectratio]{bpfig2c.epsf}
\includegraphics[width=7.5cm,keepaspectratio]{bpfig2d.epsf}
\end{center}
\caption{Typical snapshots from the canonical MC simulation of 
Figure 1(a).
The figures were created with Molscript \cite{Molscript}
and Raster3D \cite{Raster3Da,Raster3Db}.}
\label{fig2}
\end{figure}

\begin{figure}[hbtp]
\begin{center}
\includegraphics[width=7.5cm,keepaspectratio]{bpfig3a.epsf}
\includegraphics[width=7.5cm,keepaspectratio]{bpfig3b.epsf}
\includegraphics[width=7.5cm,keepaspectratio]{bpfig3c.epsf}
\includegraphics[width=7.5cm,keepaspectratio]{bpfig3d.epsf}
\end{center}
\caption{Typical snapshots from the multicanonical MC 
simulation of Figure 1(b).
The figures were created with Molscript \cite{Molscript}
and Raster3D \cite{Raster3Da,Raster3Db}.}
\label{fig3}
\end{figure}

The next examples of the systems that we studied by
multicanonical MC simulations are homo-oligomer systems.
We studied the helix-forming tendencies of three
amino-acid homo-oligomers of length 10 in gas phase
\cite{OH1,OH} and in aqueous solution (the solvent effects
are represented by
the term that is proportional to solvent-accessible
surface area) \cite{MO2}.
Three characteristic amino acids,
alanine (helix former), valine (helix indifferent), 
and glycine (helix breaker) were considered.
In Figure 4 the lowest-energy conformations obtained
both in gas phase and in aqueous solution by MUCA
simulations are shown \cite{MO2}.
The lowest-energy conformations of (Ala)$_{10}$
(Figures 2(a) and 2(b)) have six intrachain backbone 
hydrogen bonds that
characterize the $\alpha$-helix and are indeed 
completely helical.
Those of (Val)$_{10}$ are also in
almost ideal
helix state (from residue 2 to residue 9 in gas phase and
from residue 2 to residue 8 in aqueous solution).
On the other hand, those of (Gly)$_{10}$ are
not helical and rather round.

\begin{figure}[hbtp]
\begin{center}
\includegraphics[width=15.5cm,keepaspectratio]{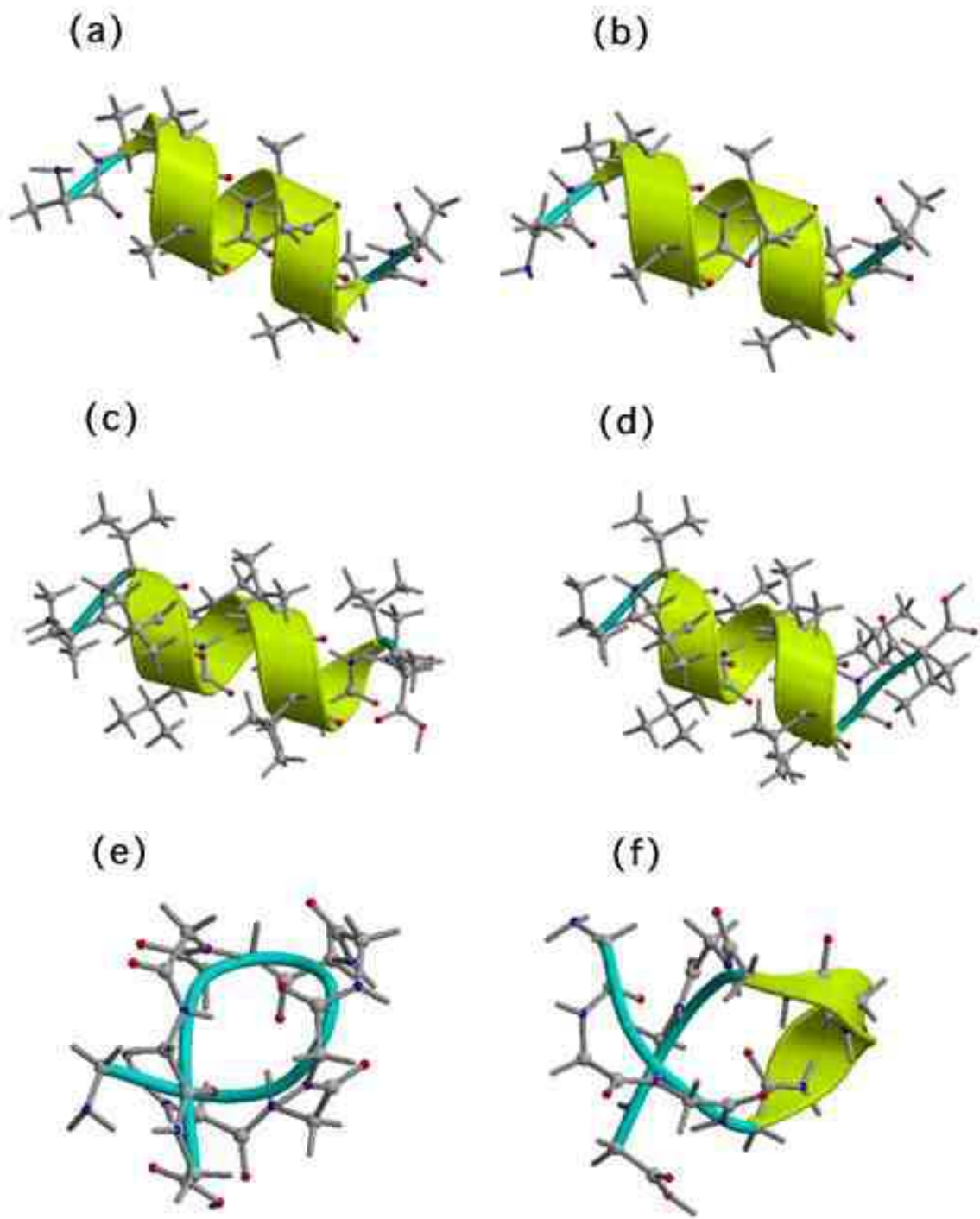}
\end{center}
\caption{The lowest-energy conformations of (Ala)$_{10}$
((a) and (b)),
(Val)$_{10}$ ((c) and (d)), and (Gly)$_{10}$ ((e) and (f))
obtained from the multicanonical MC
simulations in gas phase and in aqueous 
solution, respectively.
The figures were created with Molscript \cite{Molscript}
and Raster3D \cite{Raster3Da,Raster3Db}.}
\label{fig4}
\end{figure}

We calculated the average values of the total potential
energy and its component terms of
(Ala)$_{10}$ as a function of
temperature both in gas phase and in aqueous solution
\cite{MO2}.
The results are shown in Figure 5.
For homo-alanine in gas phase, all the conformational 
energy terms
increase monotonically as temperature increases.
The changes of each component terms are
very small except for the Lennard-Jones term,
$E_{\rm v}$, indicating that $E_{\rm v}$ plays an important role in
the folding of homo-alanine \cite{OH}.

\begin{figure}[hbtp]
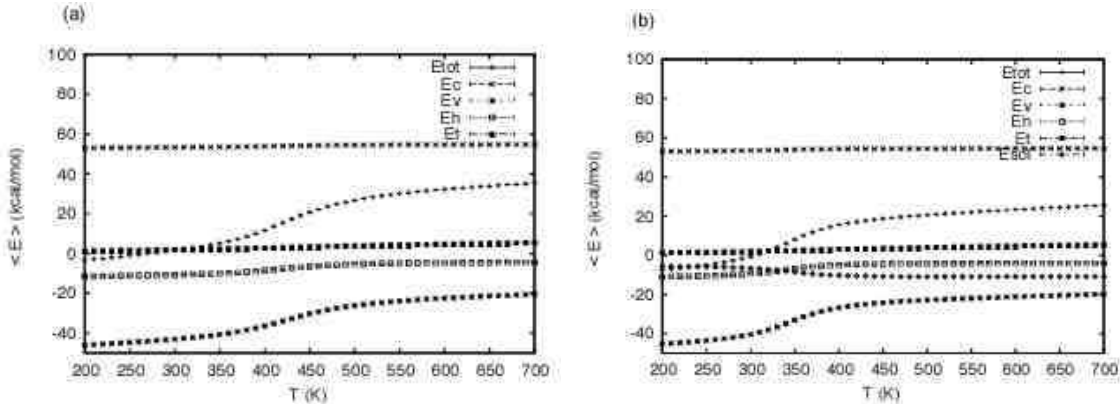

\begin{center}
\includegraphics[width=7.5cm,keepaspectratio]{bpfig5a.epsf}
\includegraphics[width=7.5cm,keepaspectratio]{bpfig5b.epsf}
\end{center}
\caption{Average of the total potential energy $E_{\rm tot}$
and averages of its component terms, electrostatic 
energy $E_{\rm c}$,
hydrogen-bond energy $E_{\rm h}$, Lennard-Jones energy $E_{\rm v}$, 
torsion energy
$E_{\rm t}$, and solvation free energy $E_{\rm sol}$ 
(only for the case
in aqueous solution) for homo-alanine as a
function of temperature $T$ (a) in gas phase and 
(b) in aqueous solution.
The values for each case were calculated from one 
multicanonical
production run of 1,000,000 MC sweeps by the
single-histogram reweighting techniques.}
\label{fig5}
\end{figure}

In aqueous solution the overall behaviors 
of the conformational energy
terms are very similar to those in gas phase.
The solvation term, on the other hand, decreases
monotonically as temperature increases.
These results imply that the solvation term favors
random-coil conformations, while the conformational terms
favor helical conformations.

The rapid changes (decrease for the solvation term and 
increase
for the rest of the terms) of all the average values 
occur at the
same temperature (around at 420 K in gas phase and 340 K 
in solvent).
We thus calculated the specific heat for 
(Ala)$_{10}$ as a function of temperature.
The specific heat here is defined by the following equation:
\begin{equation}
C(T) = {\beta}^2 \ \frac{<E_{\rm tot}^2>_T - {<E_{\rm tot}>_T}^2}{N},
\label{eqn43}
\end{equation}
where $N \ (=10)$ is the number of residues in the oligomer.
In Figure 6 we show the results.
We observe sharp peaks in the specific heat for
both environment.
The temperatures at the peak, helix-coil transition 
temperatures, are
$T_c \approx$ 420 K and 340 K in gas phase and
in aqueous solution, respectively.

\begin{figure}[hbtp]
\begin{center}
\includegraphics[width=12.0cm,keepaspectratio]{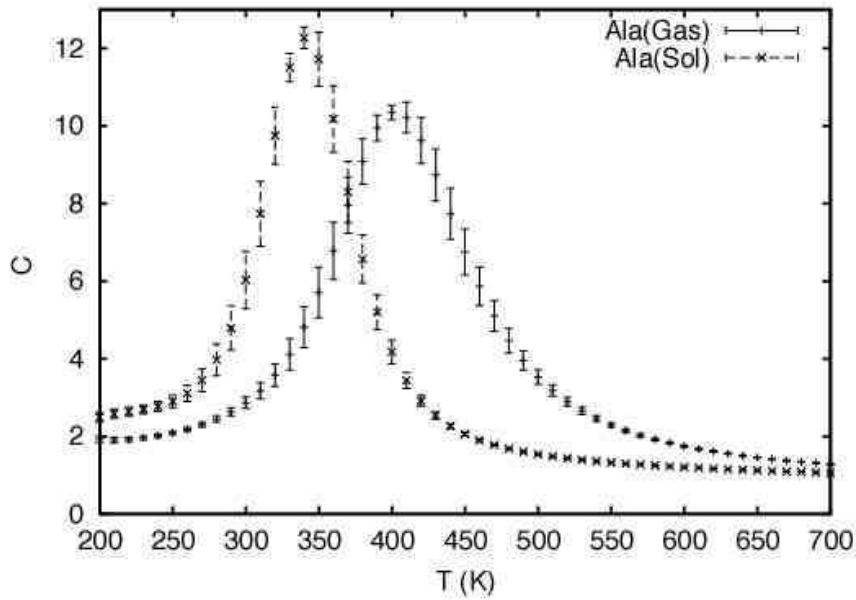}
\end{center}
\caption{Specific heat $C$ as a function of temperature $T$ for
(Ala)$_{10}$ in gas phase and in aqueous solution.
The values for each case were calculated from one 
multicanonical
production run of 1,000,000 MC sweeps by the
single-histogram reweighting techniques.}
\label{fig6}
\end{figure}

We calculated the average number of helical residues $<n>_T$
in a conformation
as a function of temperature.
In Figure 7 we show the average helicity $< n >_T$ 
as a function
of temperature for the three homo-oligomers
in aqueous solution.
The average helicity tends to decrease monotonically 
as the temperature
increases because of the increased thermal fluctuations.

At $T=200$ K, $<n>_T$ for homo-alanine is 8.
If we neglect the terminal residues, in which $\alpha$-helix
tends to be frayed, $n = 8$ corresponds to the
maximal helicity, and the conformation can be considered
completely helical.
The homo-alanine is thus in an ideal helical
structure at $T=200$ K.
Around the room temperature, the homo-alanine
is still substantially
helical
($\approx 70$ \% helicity).
This is consistent with the experimental fact that alanine 
is a 
strong helix former.
We observe that $<n>_T$ is 5 (50 \% helicity)
at the transition temperature  
obtained from the peak
in specific heat (around 340 K).
This implies that the peak
in specific heat indeed implies a helix-coil transition
between an ideal
helix and a random coil.

\begin{figure}[hbtp]
\begin{center}
\includegraphics[width=12.0cm,keepaspectratio]{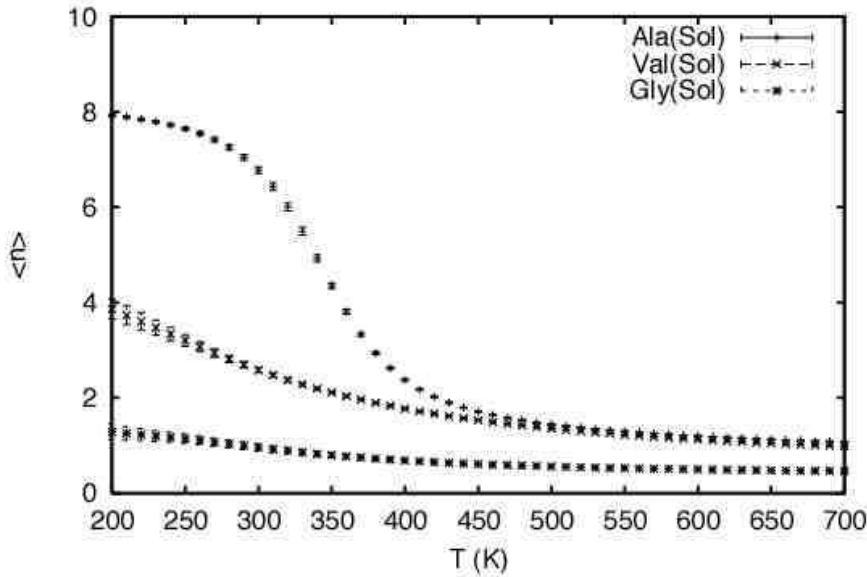}
\end{center}
\caption{Average helicity $<n>_T$ as a function of 
temperature $T$ for (Ala)$_{10}$, (Val)$_{10}$,
and (Gly)$_{10}$ in aqueous 
solution.
The values for each case were calculated from one 
multicanonical
production run of 1,000,000 MC sweeps by the
single-histogram reweighting techniques.}
\label{fig7}
\end{figure}

The next example is a penta peptide, Met-enkephalin,
whose amino-acid sequence is: Tyr-Gly-Gly-Phe-Met.
Since this is one of the simplest peptides with
biological functions, it served as a bench mark
system for many simulations.

Here, we present the latest results of a multicanonical 
MC simulation of Met-enkephalin in gas phase \cite{MHO}.
The conformations were classified into six groups of
similar structures according to their intra-chain
hydrogen bonds.  In Figure 8 we show the
lowest-energy conformations in each group identified
by the MUCA simulation.
The lowest-energy conformation of group C25 (Figure 8(a))
has two hydrogen
bonds, connecting residues 2 and 5, and forms a type 
II$^{\prime}$ $\beta$-turn.
The ECEPP/2 energy of the conformation is $-12.2$ kcal/mol, 
and this conformation
corresponds to the global-minimum-energy state of 
Met-enkephalin in gas phase.
The lowest-energy conformation of group C14 (Figure 8(b)) 
has two hydrogen bonds, connecting residues 1 and 4,
and forms a type II $\beta$-turn.
The energy is $-$11.1 kcal/mol, and this conformation 
corresponds to the
second-lowest-energy state.  Other groups correspond to
high-energy
states.

\begin{figure}[hbtp]
\begin{center}
\includegraphics[width=15.5cm,keepaspectratio]
{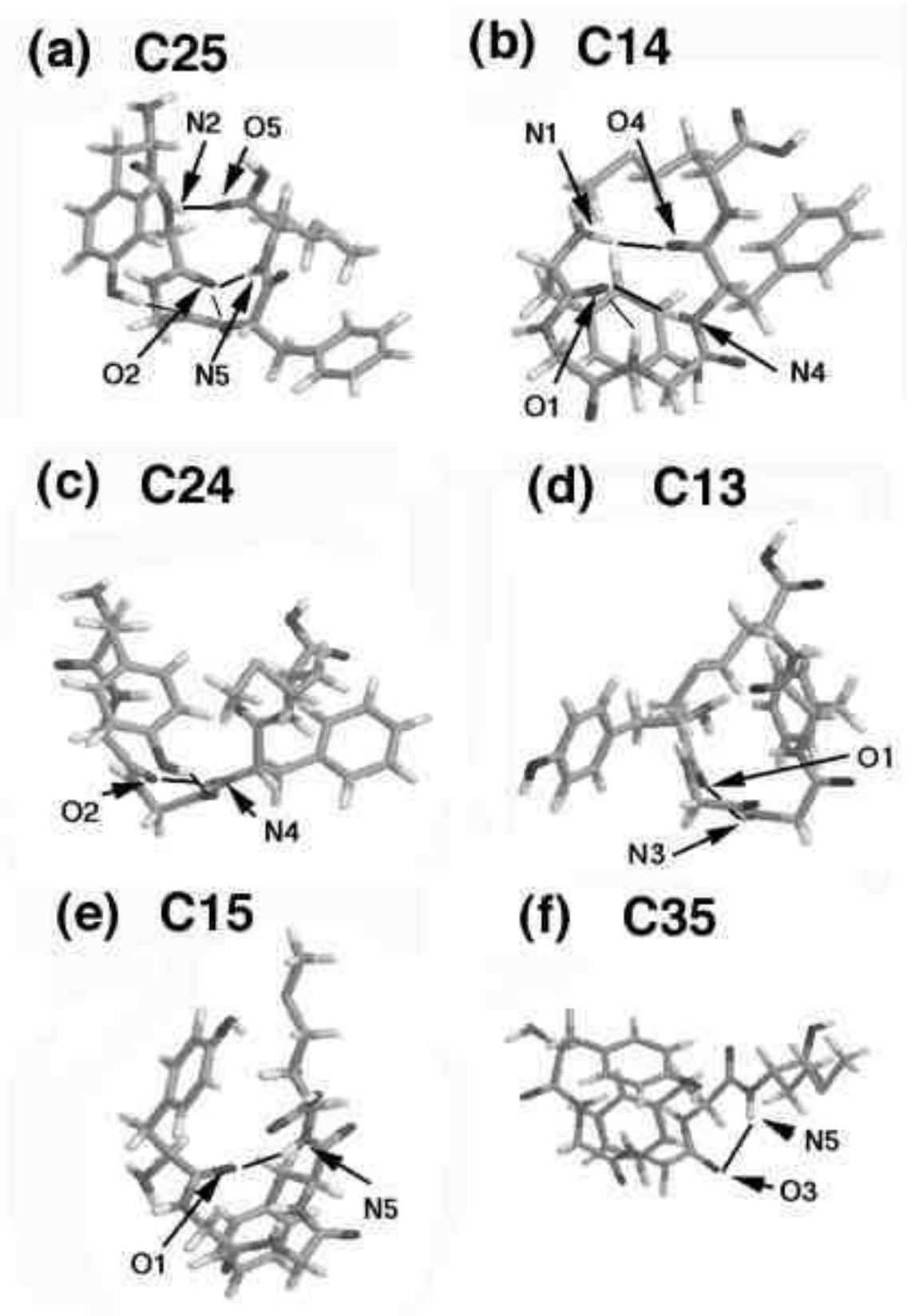}
\end{center}
\caption{The lowest-energy conformations in each group
obtained by the multicanonical MC simulation of
1,000,000 MC sweeps.
The lowest-energy conformations correspond to
groups (a) C25, (b) C14, (c) C24, (d) C13, 
(e) C15, and (f) C35.
The figures were created with RasMol \cite{RasMol}.}
\label{fig8}
\end{figure}

We now study
the distributions of conformations in these groups 
as a function of
temperature. The results are shown in Figure 9.
As can be seen in the Figure, group C25
is dominant at low temperatures. Conformations of 
group C14 start to appear
from $T \approx 100$ K.
At $T \approx 300$ K, the distributions of these 
two groups, C25 and
C14, balance ( $\approx$ 25 \% each) and constitute 
the main groups.
Above $T \approx 300$ K, the contributions of 
other groups become
non-negligible (those of group C24 and group C13 
are about 10 \% and 8 \%,
respectively, at $T=400$ K).
Note that the distribution of conformations that 
do not belong to any of the
six groups monotonically increases as the temperature 
is raised.
This is because random-coil conformations without 
any intrachain hydrogen
bonds are favored at high temperatures.

\begin{figure}[hbtp]
\begin{center}
\includegraphics[width=12.0cm,keepaspectratio]
{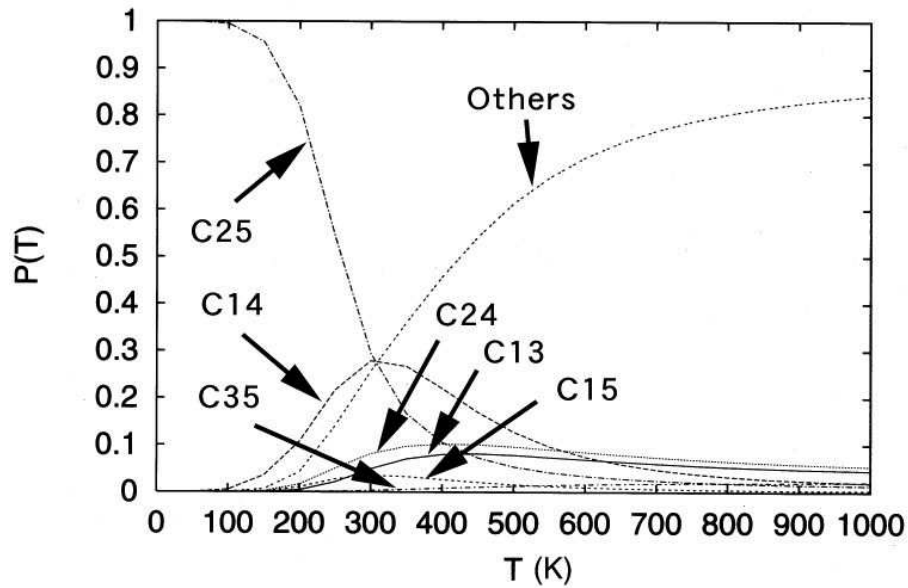}
\end{center}
\caption{The distributions of each group of similar 
strucutres
as a function of temperature.}
\label{fig9}
\end{figure}

The same peptide in gas phase was studied by
the replica-exchange MD simulation \cite{SO}.
We made an MD
simulation of $2 \times 10^6$ time steps (or, 1.0 ns)
for each replica, starting from an extended conformation. 
We used the following eight temperatures: 
700, 585, 489, 409, 342, 286, 239, and 200 K, 
which are distributed 
exponentially, following the annealing schedule of simulated 
annealing simulations \cite{KONF2}.  As is shown below, 
this choice already
gave an optimal temperature distribution.
The replica exchange was tried every 10 fs, and the data 
were stored
just before the replica exchange for later analyses.

As for expectation values of physical quantities at 
various temperatures, we used the multiple-histogram
reweighting techniques of Eqs.~(\ref{Eqn8a}) and (\ref{Eqn8b}).
We remark that for biomolecular systems the integrated
autocorrelation times, $\tau_m$, in the reweighting formulae
(see Eq.~(\ref{Eqn8a}))
can safely be set to be a constant \cite{WHAM}, and we
do so throughout the analyses in this section.

For an optimal performance of REM simulations
the acceptance ratios
of replica exchange should be
sufficiently uniform and large (say, $> 10$ \%).
In Table 1 we list these quantities.
The values are indeed uniform (all about 15 \% of
acceptance probability)
and large enough (more than 10 \%). 

\begin{table}[hbtp]
\caption{Acceptance Ratios of Replica Exchange Corresponding to
Pairs of Neighboring Temperatures}
\begin{center}
\begin{tabular}{cc} \hline
Pair of Temperatures (K) & Acceptance Ratio \\
\hline
200 $\longleftrightarrow$ 239 & 0.160 \\
239 $\longleftrightarrow$ 286 & 0.149 \\
286 $\longleftrightarrow$ 342 & 0.143 \\
342 $\longleftrightarrow$ 409 & 0.139 \\
409 $\longleftrightarrow$ 489 & 0.142 \\
489 $\longleftrightarrow$ 585 & 0.146 \\
585 $\longleftrightarrow$ 700 & 0.146 \\
\hline
\end{tabular}
\end{center}

\label{Tab1}
\end{table}

The results in Table 1 imply that one should observe a free random
walk in temperature space.  
The results for one of the replicas
are shown in Figure 10(a).  We do observe a random walk in
temperature space between the lowest and highest 
temperatures.  In Figure 10(b) the corresponding time series of
the total potential energy is shown.
We see that a random walk in potential energy space
between low and high energies is realized.
We remark that the potential
energy here is that of AMBER in Ref.~\cite{AMBER1}.
Note that there is a strong correlation between the behaviors
in Figures 10(a) and 10(b).

\begin{figure}[hbtp]
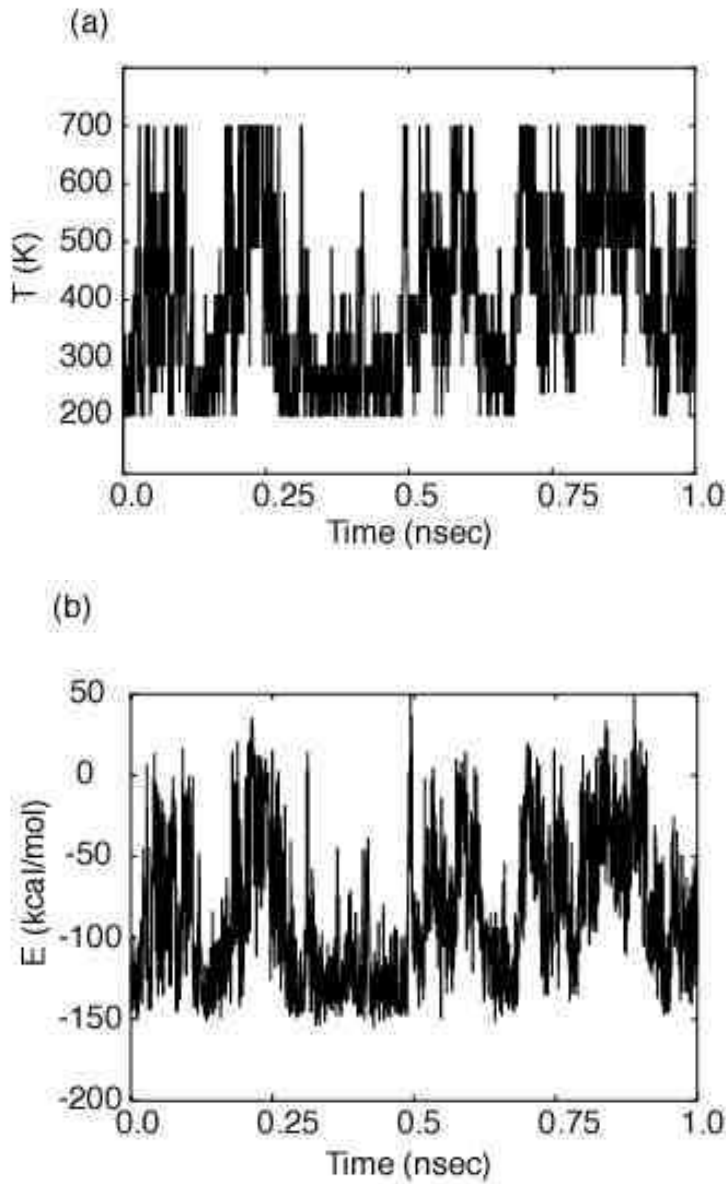

\begin{center}
\includegraphics[width=10.0cm,keepaspectratio]{bpfig10a.epsf}
\includegraphics[width=10.0cm,keepaspectratio]{bpfig10b.epsf}
\end{center}
\caption{Time series of (a) temperature exchange
and (b) the total potential energy 
for one of the replicas from a replica-exchange MD simulation
of Met-enkephalin in gas phase.}
\label{fig10}
\end{figure}

In Figure 11 the canonical probability distributions obtained
at the chosen eight temperatures from
the replica-exchange simulation are shown.  We see that there
are enough overlaps between all pairs of distributions, indicating
that there will be sufficient numbers of replica exchanges 
between pairs of replicas (see Table 1).

\begin{figure}[hbtp]
\begin{center}
\includegraphics[width=12.0cm,keepaspectratio]{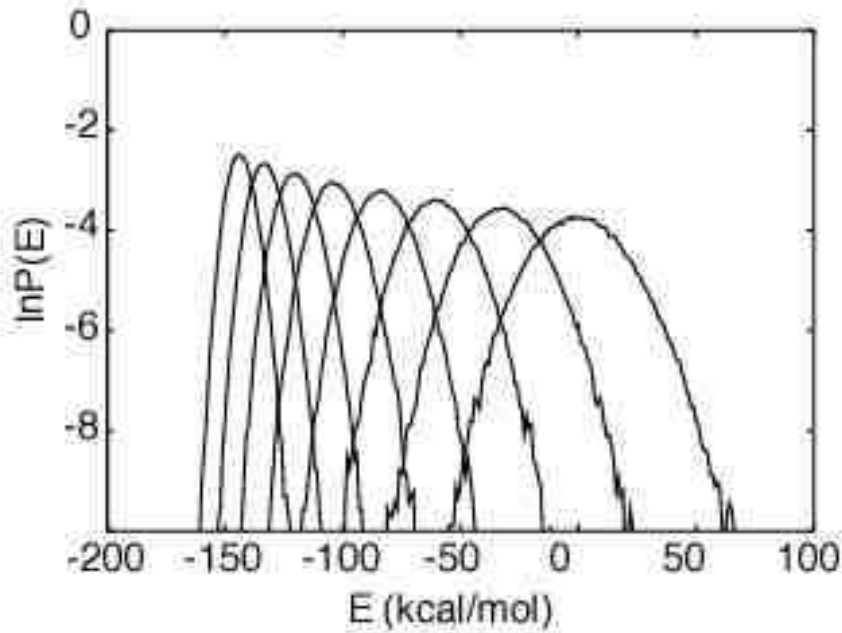}
\end{center}
\caption{The canonical probability distributions of
the total potential energy of Met-enkephalin in gas phase
obtained from
the replica-exchange MD simulation at the eight temperatures.
The distributions
correspond to the following temperatures (from left to right):
200, 239, 286, 342, 409, 489, 585, and 700 K.}
\label{fig11}
\end{figure}

We further compare the results of the replica-exchange simulation
with those of a single canonical MD simulation (of 1 ns)
at the corresponding temperatures.
In Figure 12 we compare the distributions of a pair of dihedral
angles $(\phi,\psi)$ of Gly-2 at two extreme
temperatures ($T=200$ K and 700 K).
While the results at $T=200$ K from the regular canonical simulation
are localized with only one dominant peak, those from the 
replica-exchange simulation have
several peaks (compare Figures 12(a) and 12(b)).
Hence, the replica-exchange
run samples much broader configurational space than the conventional
canonical run at low temperatures.
The results at $T=700$ K (Figures 12(c) and 12(d)), 
on the other hand,
are similar, implying that a regular canonical simulation can
give accurate thermodynamic quantities at high temperatures.

\begin{figure}[hbtp]
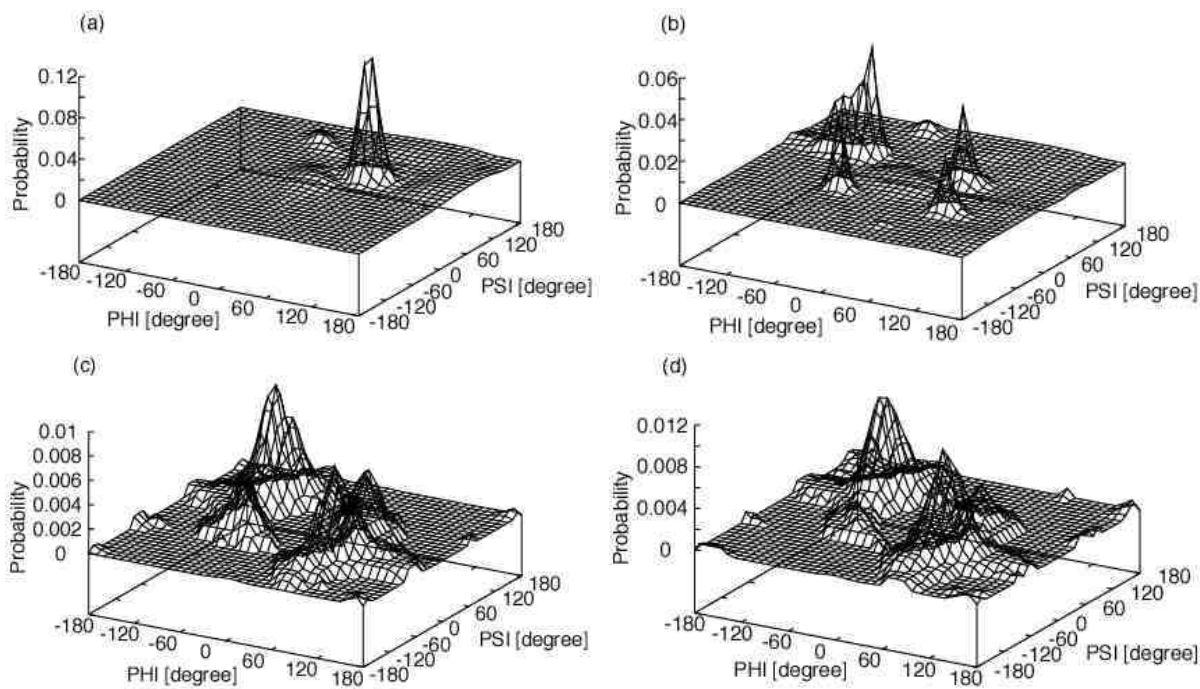

\begin{center}
\includegraphics[width=7.9cm,keepaspectratio]{bpfig12a.epsf}
\includegraphics[width=7.9cm,keepaspectratio]{bpfig12b.epsf}
\includegraphics[width=7.9cm,keepaspectratio]{bpfig12c.epsf}
\includegraphics[width=7.9cm,keepaspectratio]{bpfig12d.epsf}
\end{center}
\caption{Distributions of a pair of dihedral
angles $(\phi,\psi)$ of Gly-2 for:
(a) $T=200$ K from a regular canonical MD simulation,
(b) $T=200$ K from the replica-exchange MD simulation,
(c) $T=700$ K from a regular canonical MD simulation, and
(d) $T=700$ K from the replica-exchange MD simulation.}
\label{fig12}
\end{figure}

In Figure 13 we show the average total potential energy 
as a function of temperature.
As expected from the results of Figure 12,
we observe that the canonical simulations at low temperatures
got trapped in states of energy local minima, resulting in
the discrepancies in average values between the results from
the canonical simulations and those from the replica-exchange
simulation.

\begin{figure}[hbtp]
\begin{center}
\includegraphics[width=12.0cm,keepaspectratio]{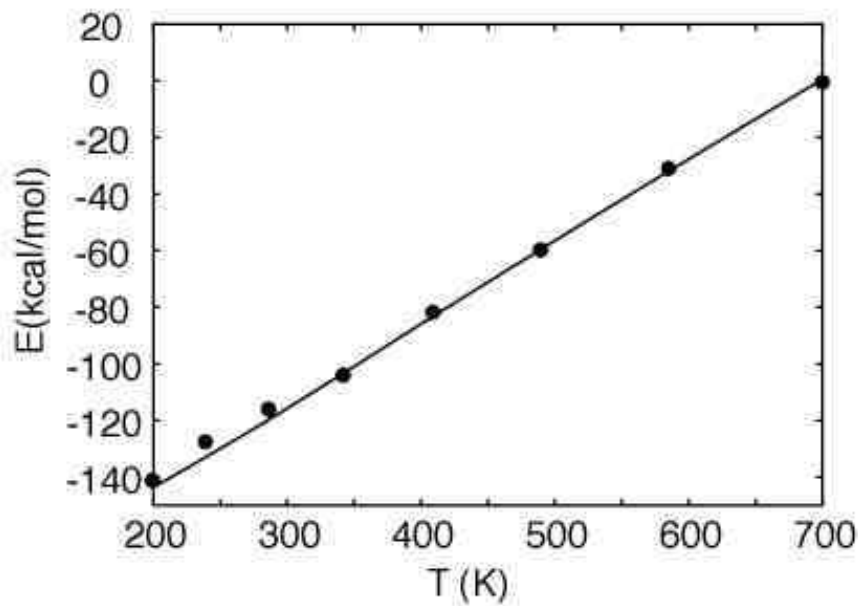}
\end{center}
\caption{Average total potential energy of Met-enkephalin
in gas phase
as a function of temperature.
The solid curve is the result from the replica-exchange
MD simulation and the dots are those of regular canonical
MD simulations.}
\label{fig13}
\end{figure}

We now present the results of MD simulations based
on replica-exchange multicanonical
algorithm and multicanonical replica-exchange method \cite{SO3}.
The Met-enkephalin in gas phase was studied again.
The potential
energy is, however, that of AMBER in Ref.~\cite{AMBER2} instead
of Ref.~\cite{AMBER1}.
In Table 2 we summarize the parameters of the simulations that
were performed.
As discussed in the previous section, REMUCA consists of
two simulations: a short REM simulation (from which the
density of states of the system, or the multicanonical weight factor,
is determined) and a subsequent
production run of MUCA simulation.
The former simulation is referred to as REM1 and the latter
as MUCA1 in Table 2.
A production run of MUCAREM simulation is referred to as
MUCAREM1 in Table 2,
and it uses the same density of states that was obtained
from REM1.
Finally, a production run of the original REM simulation
was also performed for comparison and it is referred to as
REM2 in Table 2.
The total simulation time for the three production runs
(REM2, MUCA1, and MUCAREM1) was all set equal (i.e., 5 ns).

\begin{table}[hbtp]
\caption{Summary of Parameters in REM, REMUCA, and MUCAREM Simulations}
 \begin{center}
 \begin{tabular}{cccc} \hline
   Run     & No. of Replicas, $M$ & Temperature, $T_m$ (K)
   ($m = 1, \cdots, M$) & MD Steps\\
   \hline
   REM1    & 10  & 200, 239, 286, 342, 409, & $2 \times 10^5$ \\
           &     & 489, 585, 700, 836, 1000 & \\
   REM2    & 10  & 200, 239, 286, 342, 409, & $1 \times 10^6$ \\
           &     & 489, 585, 700, 836, 1000 & \\
   MUCA1   & 1   & 1000 & $1 \times 10^7$ \\
   MUCAREM1 & 4    & 375, 525, 725, 1000     & $2.5 \times 10^6$ \\
   \hline
  \end{tabular}
 \end{center}
 \label{Tab2}
\end{table}

After the simulation of REM1 is finished, we obtained the 
density of states, $n(E)$, by the
multiple-histogram reweighting techniques of Eqs.~(\ref{Eqn8a}) 
and (\ref{Eqn8b}).
The density of states will give the average values of the
potential energy from Eq.~(\ref{eqn18}), and we found
\begin{equation}
\left\{
\begin{array}{rl}
E_1 &=~ <E>_{T_1} = -30 ~{\rm kcal/mol}~, \\
E_M &=~ <E>_{T_M} = 195 ~{\rm kcal/mol}~.
\end{array}
\right.
\label{eqn50}
\end{equation}
Then our estimate of the density of states is reliable
in the range $E_1 \le E \le E_M$.
The multicanonical potential energy ${\cal E}_{mu}^{\{0\}}(E)$
was thus determined for the three energy regions
($E < E_1$, $E_1 \le E \le E_M$, and $E > E_M$) from
Eq.~(\ref{eqn31}).
Namely, the multicanonical potential energy, $E_{mu}(E;T_0)$,
in Eq.~(\ref{eqn7}) and its derivative,
$\frac{\partial E_{mu}(E;T_0)}{\partial E}$,
were determined by fitting $\ln n(E)$ by cubic
spline functions in the energy region of
($-30 \le E \le 195$ kcal/mol) \cite{SO3}.
Here, we have set the arbitrary reference temperature to 
be $T_0 = 1000$ K.
Outside this energy region, $E_{mu}(E;T_0)$ was linearly
extrapolated as in Eq.~(\ref{eqn31}).

After determining the multicanonical weight factor,
we carried out a multicanonical MD simulation of $1 \times
10^7$ steps (or 5 ns) for data collection (MUCA1 in Table 2).
In Figure 14 the probability distribution obtained by 
MUCA1 is plotted.
It can be seen that a good
flat distribution is obtained in the energy region
$E_1 \le E \le E_M$.
In Figure 14 the canonical probability distributions that
were obtained by the reweighting techniques at
$T = T_1 = 200$ K and $T = T_M = 1000$ K are also
shown (these results are essentially identical to
one another among
MUCA1, MUCAREM1, and REM2, as discussed below).
Comparing these curves with those of MUCA1 in
the energy regions $E < E_1$
and $E > E_M$ in Figure 14, we confirm our claim in the
previous section that
MUCA1 gives canonical distributions at $T=T_1$ for
$E < E_1$ and at $T=T_M$ for $E > E_M$, whereas
it gives a multicanonical distribution for
$E_1 \le E \le E_M$.

\begin{figure}[hbtp]
\begin{center}
\includegraphics[width=12.0cm,keepaspectratio]{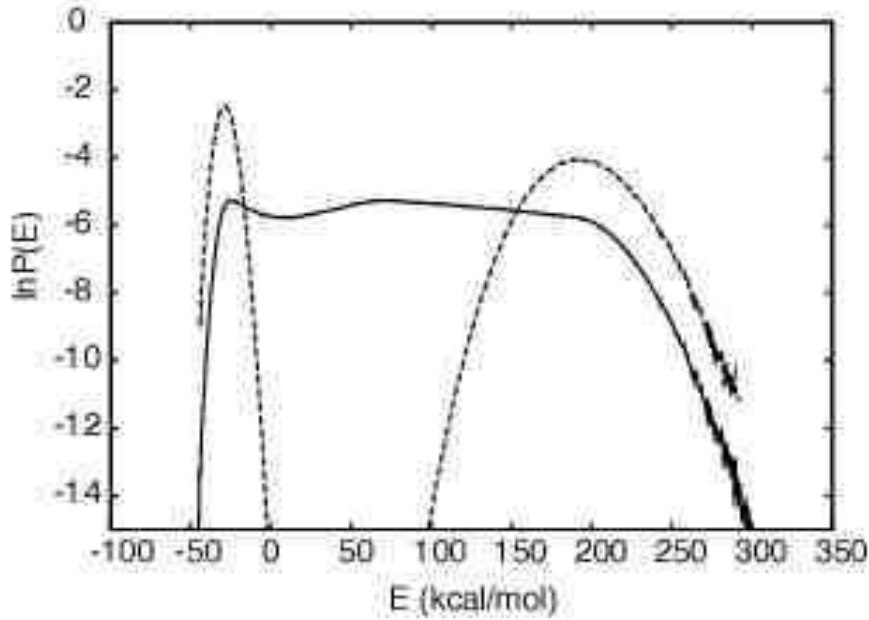}
\end{center}
\caption{Probability distribution of potential energy
        of Met-enkephalin in gas phase that was obtained from MUCA1
        (see Table 2).  The dotted curves are the probability
       distributions of the reweighted
       canonical ensemble at $T = 200$ K (left) and 1000 K (right).}
\label{fig14}
\end{figure}

In the previous
works of multicanonical simulations of Met-enkephalin in gas phase
(see, for instance, Refs.~\cite{HO,MHO}), at least several
iterations of trial simulations were required for the multicanonical
weight determination.
We emphasize that in the present case of REMUCA (REM1), only
one simulation was necessary to determine the optimal multicanonical
weight factor that can cover the energy region corresponding to
temperatures between 200 K and 1000 K.

From the density of states obtained by
REMUCA (i.e., REM1),
we prepared the multicanonical weight factors (or the multicanonical
potential energies) for the MUCAREM simulation (see
Eq.~(\ref{eqn33})).
The parameters
of MUCAREM1, such as energy bounds
$E_L^{\{m\}}$ and $E_H^{\{m\}}$ ($m=1, \cdots, {\cal M}$) are
listed in Table 3.  The choices of
$T_L^{\{m\}}$ and $T_H^{\{m\}}$ are,
in general, arbitrary, but significant overlaps between the
probability distributions of adjacent replicas are necessary.
The replica-exchange process in MUCAREM1 was tried every 200
time steps (or 100 fs). It is less frequent than in REM1 (or REM2).
This is because we
wanted to ensure a sufficient time for system relaxation.

\begin{table}[hbtp]
 \caption{Summary of Parameters in MUCAREM1}
 \begin{center}
 \begin{tabular}{cccccc} \hline
    $m$ & $T_L^{\{m\}}$ (K) & $T_H^{\{m\}}$ (K) & $T_m$ (K) 
& $E_L^{\{m\}}$ (kcal/mol) & $E_H^{\{m\}}$ (kcal/mol) \\
   \hline
    1 & 200 & 375  & 375  & $-30$ & 20 \\
    2 & 300 & 525  & 525  & $-5$  & 65 \\
    3 & 375 & 725  & 725  & 20  & 120 \\
    4 & 525 & 1000 & 1000 & 65  & 195 \\
     \hline
  \end{tabular}
 \end{center}
 \label{Tab3}
\end{table}

In Figure 15 the probability distributions of potential energy
obtained by MUCAREM1 are shown.
As expected, we observe that the
probability distributions corresponding to the temperature
$T_m$ are essentially flat for the energy region
$E_L^{\{m\}} \le E \le E_H^{\{m\}}$, are of the canonical simulation at
$T=T_L^{\{m\}}$ for $E < E_L^{\{m\}}$, and are of the
canonical simulation at $T=T_H^{\{m\}}$ for $E > E_H^{\{m\}}$
($m=1, \cdots, {\cal M}$).
As a result, each distribution in MUCAREM is much broader than those
in the conventional REM and a much smaller number of
replicas are required in MUCAREM than in REM
(${\cal M}=4$ in MUCAREM versus $M=10$ in REM).

\begin{figure}[hbtp]
\begin{center}
\includegraphics[width=12.0cm,keepaspectratio]{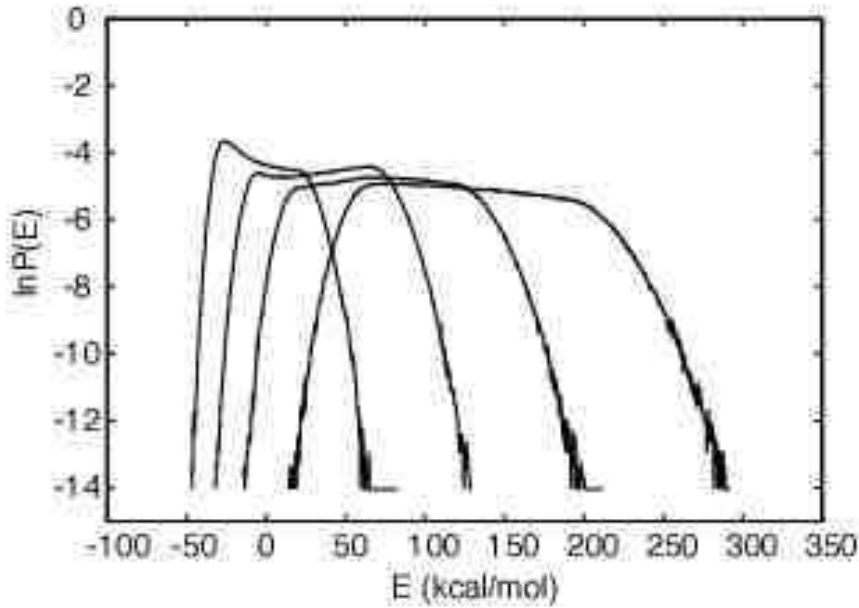}
\end{center}
\caption{Probability distributions of potential energy
       obtained from MUCAREM1 (see Tables 2 and 3).}
\label{fig15}
\end{figure}

In Figure 16 the time series of potential energy for the
first 500 ps of REM2 (a), MUCA1 (b), and MUCAREM1 (c) are plotted.
They all exhibit a random walk in potential energy space,
implying that they all perfomed properly as generalized-ensemble
algorithms.
To check the validity of the canonical-ensemble expectation values
calculated by the new algorithms, we
compare the average potential energy as a function of temperature
in Figure 17.
In REM2 and MUCAREM1 we used the multiple-histogram
techniques \cite{FS2,WHAM}, whereas the
single-histogram method \cite{FS1}
was used in MUCA1. We can see a perfect coincidence of
these quantities among REM2, MUCA1, and MUCAREM1 in Figure 17.

\begin{figure}[hbtp]
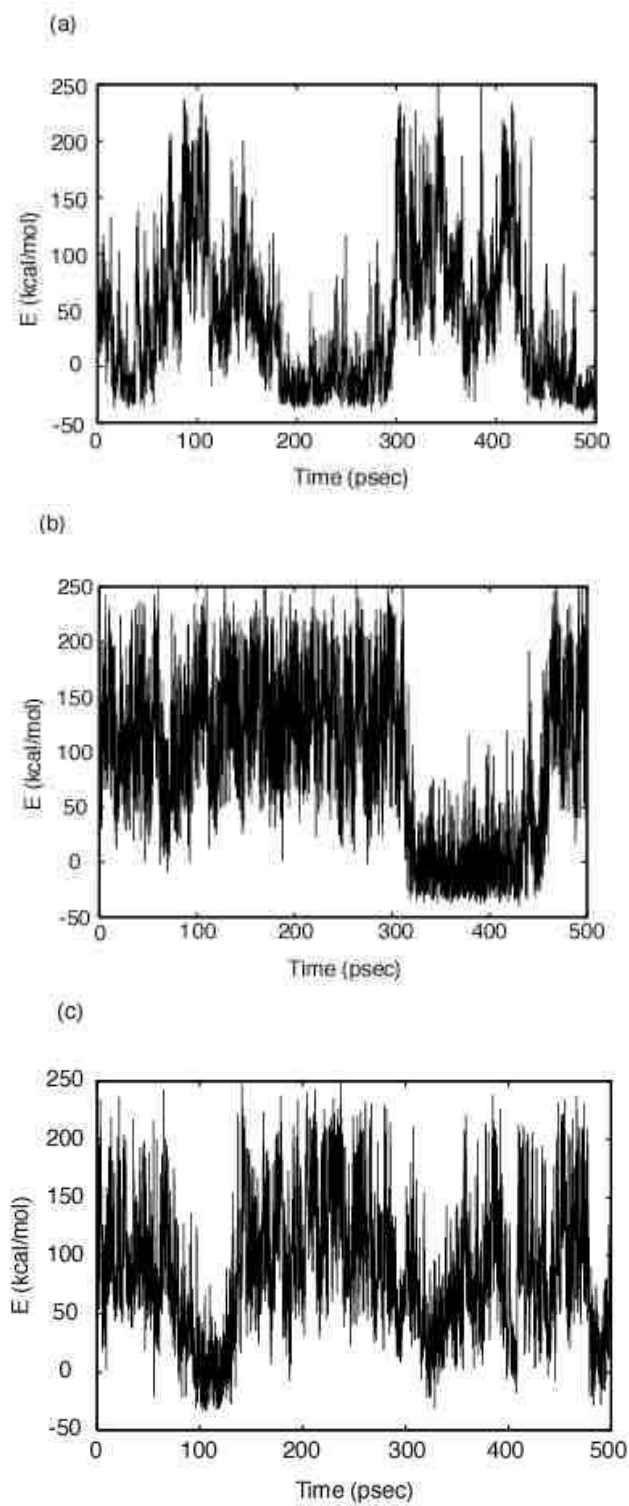

\begin{center}
\includegraphics[width=8.5cm,keepaspectratio]{bpfig16a.epsf}
\includegraphics[width=8.5cm,keepaspectratio]{bpfig16b.epsf}
\includegraphics[width=8.5cm,keepaspectratio]{bpfig16c.epsf}
\end{center}
\caption{Time series of potential energy of Met-enkephalin
in gas phase for one of the
       replicas in (a) REM2, (b) MUCA1, and (c) MUCAREM1
               (see Tables 2 and 3 for the parameters
               of the simulations).}
\label{fig16}
\end{figure}

\begin{figure}[hbtp]
\begin{center}
\includegraphics[width=10.0cm,keepaspectratio]{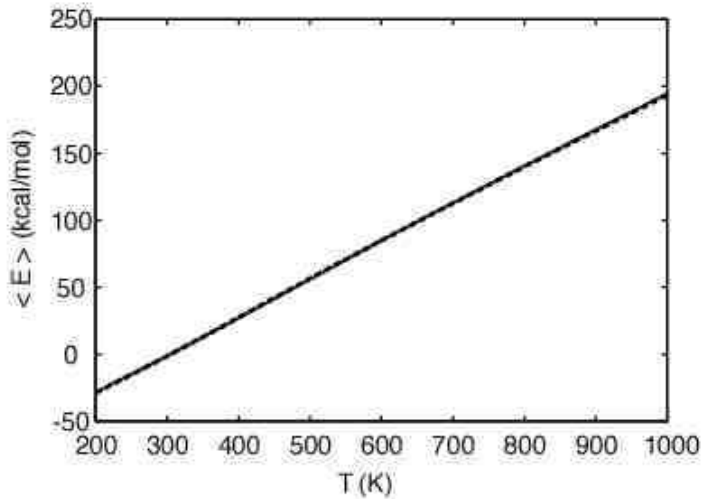}
\end{center}
\caption{The average potential energy of Met-enkephalin in
       gas phase as a function of temperature.
       The solid, dotted, and dashed curves are obtained from REM2,
       MUCA1, and MUCAREM1, respectively
       (see Tables 2 and 3 for the parameters of the simulations).}
\label{fig17}
\end{figure}

We now present the results of a replica-exchange
simulated tempering MC simulation of Met-enkephalin
in gas phase \cite{MO4}.  The potential
energy is again
that of ECEPP/2 \cite{ECEP1}--\cite{ECEP3}.
In Table 4 we summarize the parameters of the simulations that
were performed.
As described in the previous section, REST consists of
two simulations: a short REM simulation (from which the
dimensionless Helmholtz free energy, or the simulated tempering
weight factor, is determined) and a subsequent
ST production run.
The former simulation is referred to as REM1 and the latter
as ST1 in Table 4.
In REM1 there exist
8 replicas with 8 different temperatures ($M=8$), ranging from
50 K to 1000 K as listed in Table 4 (i.e., $T_1 = 50$ K
and $T_M = T_{8} = 1000$ K).
The same set of temperatures were also used in ST1.
The temperatures were distributed exponentially between
$T_1$ and $T_M$, following the optimal
distribution found in the previous simulated annealing
schedule \cite{KONF2}, simulated tempering run \cite{HO96b},
and replica-exchange simulation \cite{SO}.
After estimating the weight factor, we made a ST production
run of $10^6$ MC sweeps (ST1).
In REM1 and ST1, a replica exchange and a temperature update,
respectively, were tried every 10 MC sweeps.

\begin{table}[hbtp]
 \caption{Summary of Parameters in REST Simulations}
 \begin{center}
 \begin{tabular}{cccc} \hline
   Run     & No. of Replicas, $M$ & Temperature, $T_m$ (K) ($m = 1, \cdots, M$) 
   & MC Sweeps \\
   \hline
   REM1    & 8  & 50, 77, 118, 181, 277, 425, 652, 1000 & $5 \times 10^4$ \\
   ST1    & 1  & 50, 77, 118, 181, 277, 425, 652, 1000 & $1 \times 10^6$ \\
    \hline
  \end{tabular}
 \end{center}
 \label{Tab4}
\end{table}

We first check whether the replica-exchange simulation
of REM1 indeed performed properly.
For an optimal performance of REM the acceptance ratios
of replica exchange should be
sufficiently uniform and large (say, $> 10$ \%).
In Table 5 we list these quantities.
It is clear that both points are met in the sense that
they are of the same order (the values vary
between 10 \% and 40 \%).

\begin{table}[hbtp]
 \caption{Acceptance Ratios of Replica Exchange in REM1
 of Table 4}
 \begin{center}
 \begin{tabular}{cc} \hline
   Pair of Temperatures (K) & Acceptance Ratio \\
   \hline
    50~~  $\longleftrightarrow$   ~~77 & 0.30 \\
    77~~  $\longleftrightarrow$   ~~118 & 0.27 \\
    118~~  $\longleftrightarrow$   ~~181 & 0.22 \\
    181~~  $\longleftrightarrow$   ~~277 & 0.17 \\
    277~~  $\longleftrightarrow$   ~~425 & 0.10 \\
    425~~  $\longleftrightarrow$   ~~652 & 0.27 \\
    652~~  $\longleftrightarrow$   ~~1000 & 0.40 \\
     \hline
  \end{tabular}
 \end{center}
 \label{Tab5}
\end{table}

After determining the simulated tempering weight factor,
we carried out a long ST simulation
for data collection (ST1 in Table 4).
In Figure 18 the time series of temperature and
potential energy from ST1 are plotted.
In Figure 18(a) we observe a random walk in
temperature space between the lowest and highest
temperatures.  In Figure 18(b) the corresponding random
walk of the total potential energy
between low and high energies is observed.
Note that there is a strong correlation between the behaviors
in Figures 18(a) and 18(b), as there should.
It is known from our previous works that the
global-minimum-energy conformation for Met-enkephalin
in gas phase
has the ECEPP/2 energy value
of $-12.2$ kcal/mol \cite{HO94,MHO}.
Hence, the random walk in Figure 12(b) indeed visited
the global-minimum region many times.  It also visited
high energy regions, judging from the fact that
the average potential energy
is around 15 kcal/mol at $T = 1000$ K \cite{HO,MHO} (see also Figure 19 below).

\begin{figure}[hbtp]
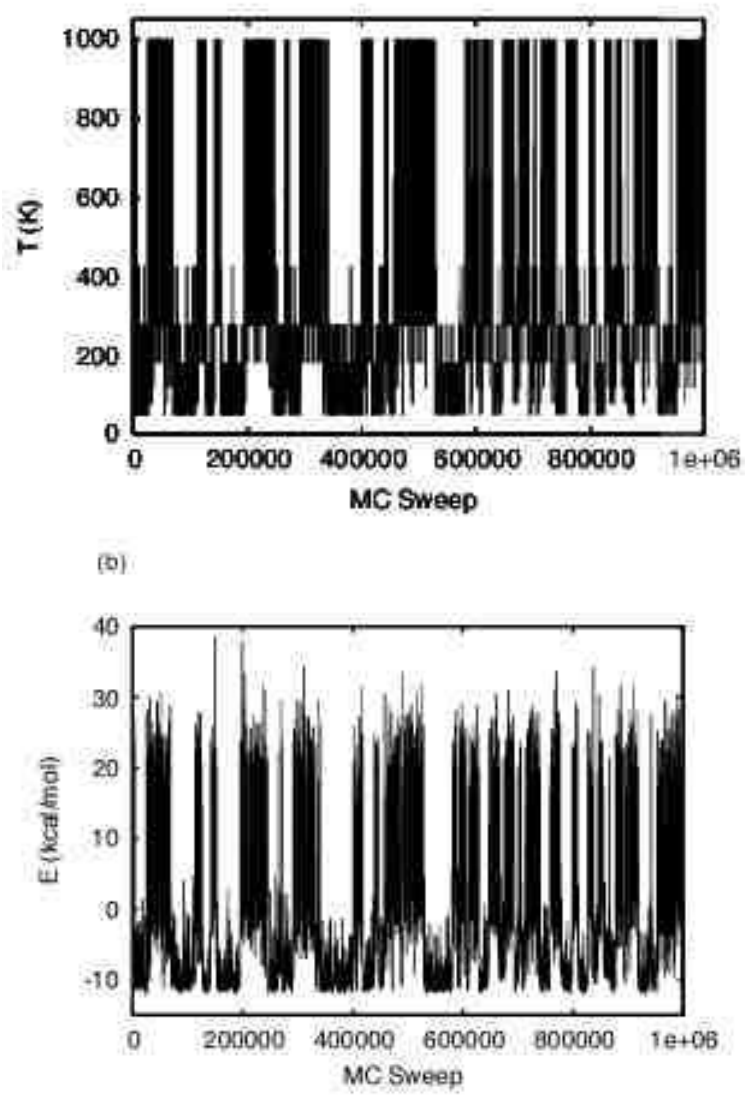

\begin{center}
\includegraphics[width=10.0cm,keepaspectratio]{bpfig18a.epsf}
\includegraphics[width=10.0cm,keepaspectratio]{bpfig18b.epsf}
\end{center}
\caption{Time series of (a) temperature and (b) potential energy
       in ST1 (see Table 4 for the parameters
               of the simulation).}
\label{fig18}
\end{figure}

For an optimal performance of ST, the acceptance ratios
of temperature update should be
sufficiently uniform and large.
In Table 6 we list these quantities.
It is clear that both points are met (the values vary
between 26 \% and 57 \%);
we find that the present ST run (ST1)
indeed properly performed.
We remark that the acceptance ratios in Table 6 are
significantly larger and more uniform than those in 
Table 5, suggesting
that ST runs can sample
the configurational space more effectively
than REM runs, provided the optimal weight factor is 
obtained.

\begin{table}[hbtp]
 \caption{Acceptance Ratios of Temperature Update in ST1}
 \begin{center}
 \begin{tabular}{cc} \hline
   Pair of Temperatures (K) & Acceptance Ratio \\
   \hline
    50~~  $\longrightarrow$   ~~77 & 0.47 \\
    77~~  $\longrightarrow$   ~~50 & 0.47 \\
    77~~  $\longrightarrow$   ~~118 & 0.43 \\
    118~~  $\longrightarrow$   ~~77 & 0.43 \\
    118~~  $\longrightarrow$   ~~181 & 0.37 \\
    181~~  $\longrightarrow$   ~~118 & 0.42 \\
    181~~  $\longrightarrow$   ~~277 & 0.29 \\
    277~~  $\longrightarrow$   ~~181 & 0.29 \\
    277~~  $\longrightarrow$   ~~425 & 0.30 \\
    425~~  $\longrightarrow$   ~~277 & 0.26 \\
    425~~  $\longrightarrow$   ~~652 & 0.43 \\
    652~~  $\longrightarrow$   ~~425 & 0.42 \\
    652~~  $\longrightarrow$   ~~1000 & 0.57 \\
    1000~~  $\longrightarrow$   ~~652 & 0.56 \\
     \hline
  \end{tabular}
 \end{center}
 \label{Tab6}
\end{table}

We remark that the details of
Monte Carlo versions of REMUCA and MUCAREM
have also been
worked out and tested with Met-enkephalin in gas phase
\cite{MSO}.  Here in Figure 19, we just show the 
average ECEPP/2
potential energy as a function of temperature that
was calculated from the four generalized-ensemble
algorithms, MUCA, REMUCA, MUCAREM, and REST \cite{MSO}.
The results are in good agreement.

\begin{figure}[hbtp]
\begin{center}
\includegraphics[width=10.0cm,keepaspectratio]{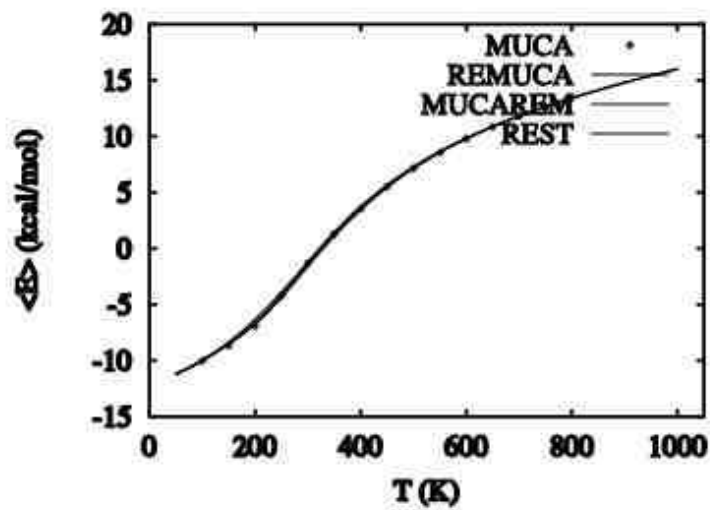}
\end{center}
\caption{The average potential energy of Met-enkephalin
  in gas phase as a function of temperature.
  The results from the four generalized-ensemble algorithms,
  MUCA, REMUCA, MUCAREM, and REST, are superimposed.}
\label{fig19}
\end{figure}

We have so far presented the results of generalized-ensemble
simulations of Met-enkephalin in gas phase.
However, peptides and proteins are usually in aqueous
solution.  We therefore want to incorporate rigorous
solvation effects in our simulations in order to
compare with experiments.

Our first example with rigorous solvent effects is
a multicanonical MC simulation, where the solvation
term was included by the RISM theory \cite{MO3}.
While low-energy conformations of Met-enkephalin
in gas phase are compact and form
$\beta$-turn structures \cite{MHO}, it turned 
out that those in aqueous solution are extended.
In Figure 20 we show the lowest-energy conformations 
of Met-enkephalin obtained during the 
multicanonical MC simulation with RISM theory incorporated
\cite{MO3}.  They
exhibit characteristics of almost fully extended
backbone structure with large side-chain fluctuations.
The results
are in accord with the observations in 
NMR experiments, which also suggest
extended conformations \cite{EnkNMR}.
 
\begin{figure}[hbtp]
\begin{center}
\includegraphics[width=8.5cm,keepaspectratio]{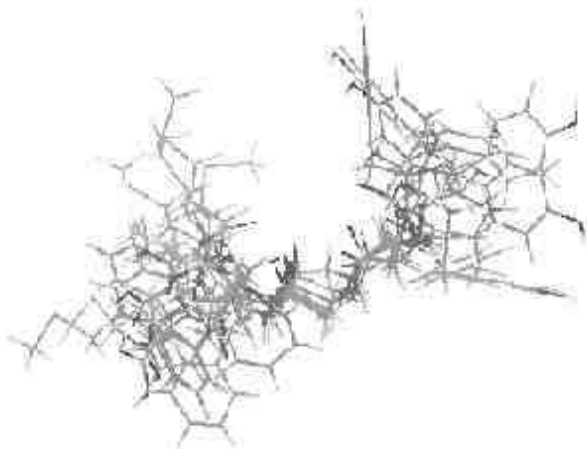}
\end{center}
\caption{Superposition of eight representative 
low-energy conformations
of Met-enkephalin
obtained by the multicanonical MC simulation 
in aqueous solution based on RISM.
The figure was created with RasMol \cite{RasMol}.}
\label{fig20}
\end{figure}

We also calculated an average
of the end-to-end distance of Met-enkephalin
as a function of temperature.
The results in aqueous solution (the present simulation)
and in the gas phase (a previous simulation  \cite{MHO}) are
compared in Figure 21.
The end-to-end distance in aqueous solution at all 
temperatures varies little (around
12 \AA);
the conformations are extended in the entire temperature range.
On the other hand, in the gas phase, the end-to-end distance is
small at low temperatures due to intrachain
hydrogen bonds, while the distance is large at high temperatures,
 because these
intrachain hydrogen bonds are broken.
 
\begin{figure}[hbtp]
\begin{center}
\includegraphics[width=12.0cm,keepaspectratio]{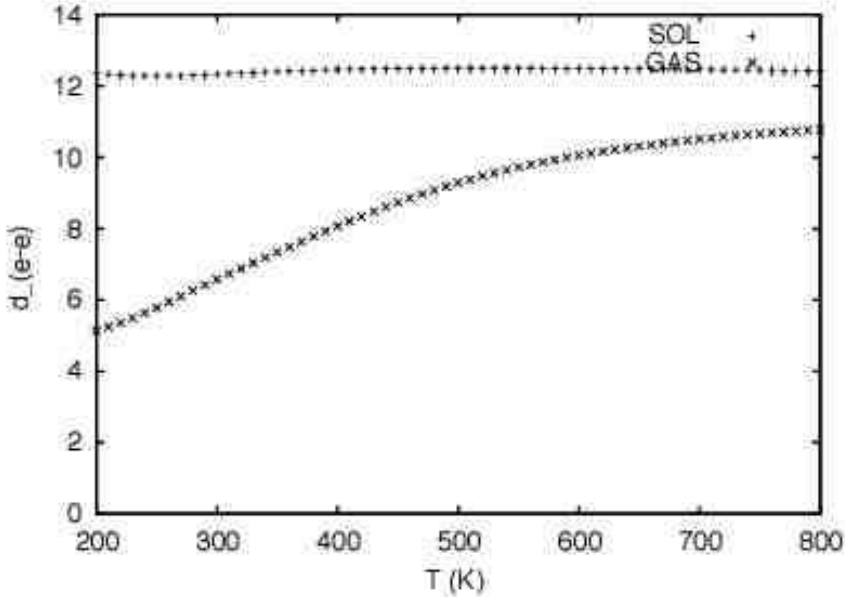}
\end{center}
\caption{Average end-to-end distance of Met-enkephalin in aqueous solution (SOL) and in gas phase (GAS)
as a function
of temperature. Here, the end-to-end distance is defined as the
distance between
the nitrogen atom at the N terminus and
the oxygen atom at the C terminus.}
\label{fig21}
\end{figure}

The same peptide was also studied by MD simulations
of replica-exchange and other generalized-ensemble
simulations in aqueous solution based on TIP3P
water model \cite{SO4}.
Two AMBER force fields \cite{AMBER2,AMBER3} were
used.  The number of water molecules was 526
and they were placed in a sphere of radius of
16 \AA.  The initial configuration is shown
in Figure 22.
 
\begin{figure}[hbtp]
\begin{center}
\includegraphics[width=10.0cm,keepaspectratio]{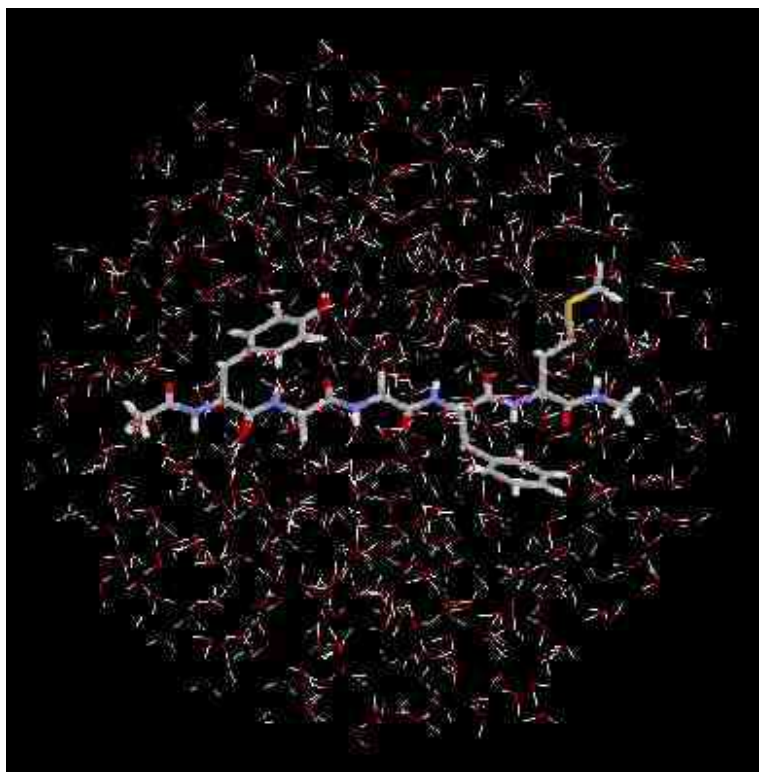}
\end{center}
\caption{Initial configuration of replica-exchange
MD simulations of Met-enkephalin in aqueous solution
with 526 TIP3P water molecules.}
\label{fig22}
\end{figure}

In Figure 23 the canonical probability distributions obtained
at the 24 temperatures from
the replica-exchange simulation are shown.  
We see that there
are enough overlaps between all pairs of distributions, 
indicating
that there will be sufficient numbers of replica exchanges 
between pairs of replicas.
The corresponding time series of the total potential
energy for one of the replicas is shown in Figure 24.
We do observe a random walk in potential energy space,
which covers an energy range of as much as 2,000 
kcal/mol.

\begin{figure}[hbtp]
\begin{center}
\includegraphics[width=12.0cm,keepaspectratio]{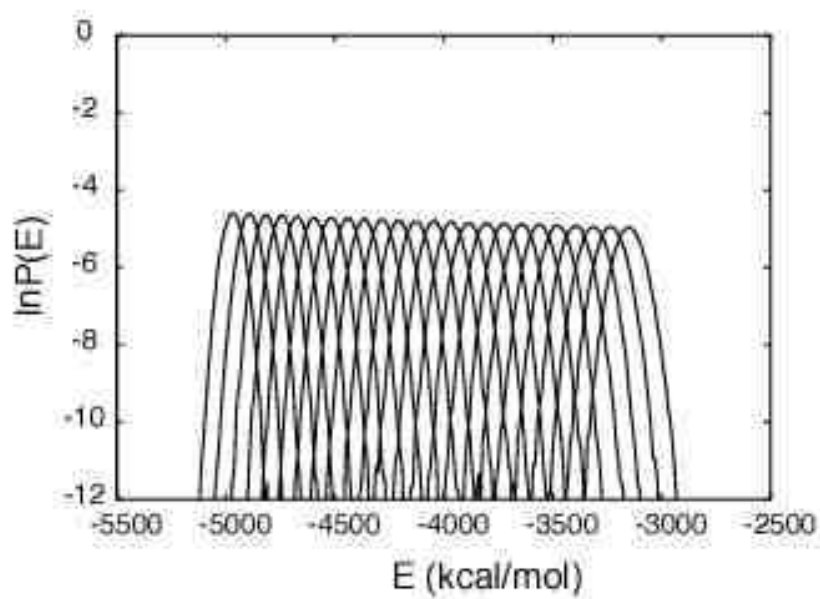}
\end{center}
\caption{The canonical probability distributions of
the total potential energy of Met-enkephalin
in aqueous solution obtained from
the replica-exchange MD simulation at the 24 temperatures.}
\label{fig23}
\end{figure}

\begin{figure}[hbtp]
\begin{center}
\includegraphics[width=12.0cm,keepaspectratio]{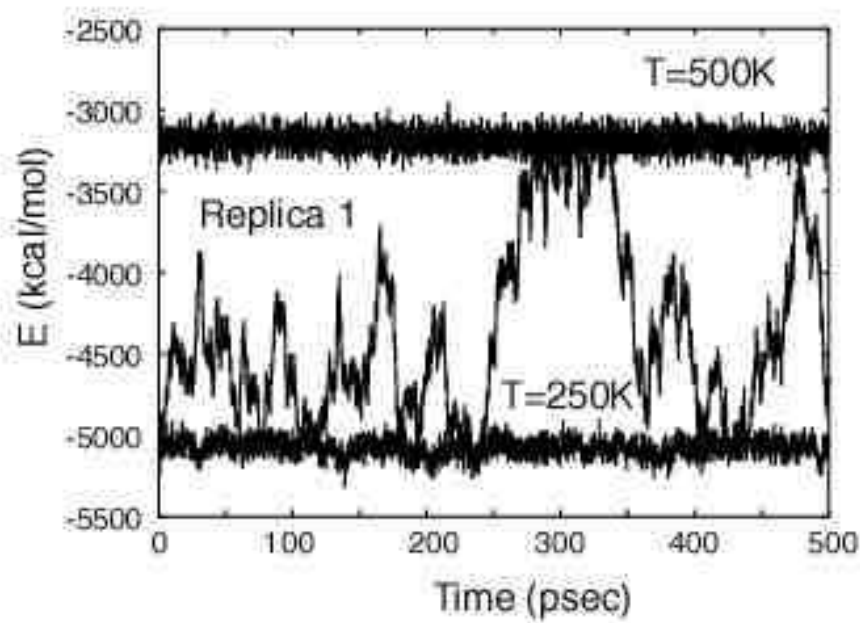}
\end{center}
\caption{Time series of
the total potential energy of Met-enkephalin
in aqueous solution obtained for one of the replicas
from
the replica-exchange MD simulation.  
Corresponding times series in the canonical ensemble
at temperatures 250 K and 500 K are also shown.}
\label{fig24}
\end{figure}

Finally, the average end-to-end distance as a function
of temperature was calculated by the multiple-histogram
reweighting techniques of Eqs.~(\ref{Eqn8a}) and 
(\ref{Eqn8b}).
The results both in gas phase and in aqueous solution
are shown in Figure 25.  The results
are in good agreement with those of ECEPP/2 energy 
plus RISM solvation theory \cite{MO3} in the sense
that Met-enkephalin is compact at low temperatures
and extended at high temperatures in gas phase and
extended in the entire temperature range
in aqueous solution (compare Figures 21 and 25).
 
\begin{figure}[hbtp]
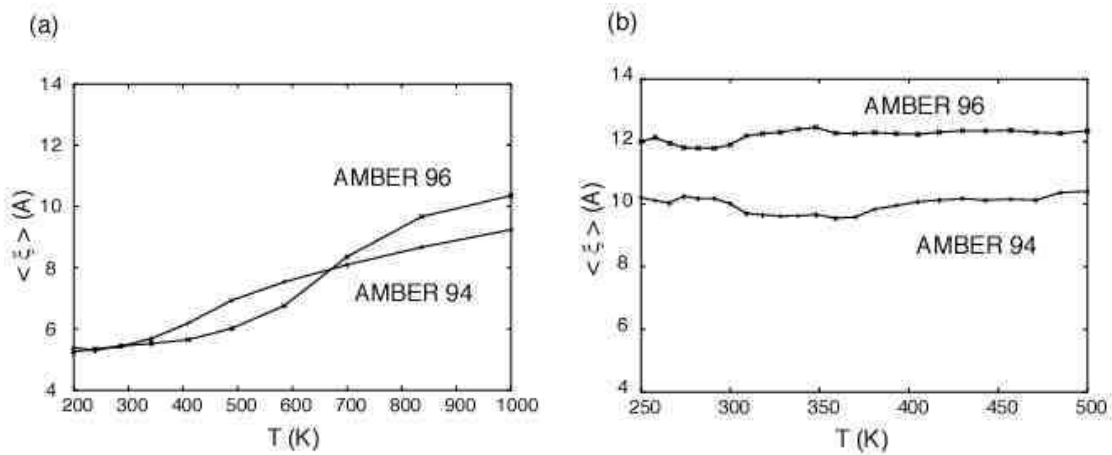

\begin{center}
\includegraphics[width=7.5cm,keepaspectratio]{bpfig25a.epsf}
\includegraphics[width=7.5cm,keepaspectratio]{bpfig25b.epsf}
\end{center}
\caption{Average end-to-end distance of Met-enkephalin 
(a) in gas phase and (b) in aqueous solution 
as a function
of temperature. }
\label{fig25}
\end{figure}

\section{CONCLUSIONS}

In this article we have reviewed uses of generalized-ensemble
algorithms in molecular simulations of biomolecules.
A simulation in generalized ensemble realizes a random
walk in potential energy space, alleviating the
multiple-minima problem that is a common difficulty
in simulations of complex systems with many degrees
of freedom.

Detailed formulations of the three well-known
generalized-ensemble algorithms, namely,
multicaonical algorithm (MUCA),
simulated tempering (ST), and replica-exchange method (REM), 
were given.
We then introduced three new generalized-ensemble
algorithms that combine the merits of the above three methods, 
which we refer to as replica-exchange multicanonical
algorithm (REMUCA), replica-exchange simulated tempering (REST),
and multicanonical replica-exchange method (MUCAREM).

With these new methods available,
we believe that we now have working simulation algorithms
which we can use for conformational predictions of
peptides and proteins from the first principles,
using the information of their amino-acid
sequence only.  It is now high time that we addressed
the question of the validity of the standard potential
energy functions such as AMBER, CHARMM, GROMOS, ECEPP, etc.
For this purpose, conventional simulations in the canonical
ensemble are of little use because they
will necessarily get trapped in states of 
local-minmum-energy states.
It is therefore essential to use generalized-ensemble
algorithms in order to test and
develop accurate potential
energy functions for biomolecular systems.
Some preliminary results of comparisons of different versions
of AMBER force fields were given in the present
article.  We remark that more detailed analyses that
compare different versions of AMBER by multicanonical
MD simulations already exist \cite{Ono}.
Likewise, the validity of solvation theories should also
be tested.  For this, RISM theory \cite{RISM1}--\cite{RISM3}
can be very useful.  For instance, we have successfully
given a molecular mechanism of secondary structural transitions
in peptides due to addition of alcohol to solvent \cite{KINO2},
which is very difficult to attain by regular molecular
simulations.

\vspace{0.5cm}
\noindent
{\bf Acknowledgements}: \\
Our simulations were performed on the Hitachi and other
computers at the Research Center for Computational
Science, Okazaki National Research Institutes.
This work is supported, in part, by a grant from 
the Research for the Future Program of the Japan Society for the 
Promotion of Science (JSPS-RFTF98P01101).\\



\noindent
\begin{thebibliography}{(00)}

\bibitem{FS1} Ferrenberg, A.M. \& Swendsen, R.H. (1988)
{\it Phys. Rev. Lett.} {\bf 61}, 2635--2638; {\it ibid.}
(1989) {\bf 63}, 1658.
\bibitem{FS2} Ferrenberg, A.M. \& Swendsen, R.H. (1989)
{\it Phys.  Rev.  Lett.} {\bf 63}, 1195--1198.
\bibitem{WHAM} Kumar, S., Bouzida, D., Swendsen, R.H., Kollman, P.A.
\& Rosenberg, J.M. (1992) {\it J. Comput. Chem.}
  {\bf 13}, 1011--1021.

\bibitem{MUCA1} Berg, B.A. \& Neuhaus, T. (1991) {\it Phys. Lett.} {\bf 
B267},
  249--253.
\bibitem{MUCA2} Berg, B.A. \& Neuhaus, T. (1992) {\it Phys. Rev. Lett.} 
{\bf 68}, 9--12.
\bibitem{MUCArev} Berg, B.A. (2000) 
{\it Fields Institute Communications} {\bf 26}, 1--24; also see
cond-mat/9909236.
\bibitem{Lee} Lee, J. (1993) {\it Phys. Rev. Lett.} {\bf 71}, 211--214;
 {\it ibid.} {\bf 71}, 2353.
\bibitem{BK} Bartels, C. \& M. Karplus, M. (1998)
 {\it J. Phys. Chem. B} {\bf 102}, 865--880.
\bibitem{BHO} Berg, B.A., Hansmann, U.H.E. \& Okamoto, Y. (1995) 
 {\it J. Phys. Chem.} {\bf 99}, 2236--2237.
\bibitem{MUCA3} Berg, B.A. \& Celik, T. (1992) {\it Phys. Rev. Lett.} 
{\bf 69}, 2292--2295.
\bibitem{MUCA4} Berg, B.A., Celik, T. \& Hansmann, U.H.E. (1993) 
{\it Europhys. Lett.} {\bf 22}, 63--68.
\bibitem{MUCA5} Berg, B.A. \& Janke, W. (1998) {\it Phys. Rev. Lett.} 
{\bf 80}, 4771--4774.
\bibitem{MUCA6} Hatano, N. \& Gubernatis, J.E. (2000)
{\it Prog. Theor. Phys. (Suppl.)} {\bf 138}, 442--447. 
\bibitem{HO} Hansmann, U.H.E. \& Okamoto, Y. (1993)
{\it J. Comput. Chem.} {\bf 14}, 1333--1338.
\bibitem{RevO} Okamoto, Y. (1998) {\it Recent Res. Devel. in Pure} \&
        {\it Applied Chem.} {\bf 2}, 1--23.
\bibitem{RevHO1} Hansmann, U.H.E. \& Okamoto, Y. (1999)
in {\it Annual Reviews of Computational Physics VI}, Stauffer, D., Ed.,
World Scientific, Singapore, pp. 129--157.
\bibitem{RevHO2} Hansmann, U.H.E. \& Okamoto, Y. (1999)
{\it Curr. Opin. Struct. Biol.} {\bf 9}, 177--183.
\bibitem{HO94} Hansmann, U.H.E. \& Okamoto, Y. (1994)
 {\it Physica} {\bf A212}, 415--437.
\bibitem{HSch} Hao, M.H. \& Scheraga, H.A. (1994)
        {\it J. Phys. Chem.} {\bf 98}, 4940--4948.
\bibitem{OH1} Okamoto, Y., Hansmann, U.H.E. \& Nakazawa, T. (1995)
   {\it Chem. Lett.} {\bf 1995}, 391--392. 
\bibitem{OH} Okamoto, Y. \& Hansmann, U.H.E. (1995)
   {\it J. Phys. Chem.} 
   {\bf 99}, 11276--11287.

\bibitem{KI95} Kidera, A. (1995) 
{\it Proc. Natl. Acad. Sci. U.S.A.} {\bf 92}, 9886--9889.
\bibitem{KGS} Kolinski, A., Galazka, W. \& Skolnick, J. (1996)
{\it Proteins} {\bf 26}, 271--287.
\bibitem{UT} Urakami, N. \& Takasu, M. (1996)
{\it J. Phys. Soc. Jpn.} {\bf 65}, 2694--2699.
\bibitem{KPV} Kumar, S., Payne, P. \& V{\' a}squez, M. (1996) 
{\it J. Comput. Chem.} {\bf 17}, 1269--1275.

\bibitem{HOE96} Hansmann, U.H.E., Okamoto, Y. \& Eisenmenger, F. (1996)
{\it Chem. Phys. Lett.} {\bf 259}, 321--330.

\bibitem{NNK} Nakajima, N., Nakamura, H. \& Kidera, A. (1997)
{\it J. Phys. Chem. B} {\bf 101}, 817--824.
\bibitem{HE} Eisenmenger, F. \& Hansmann, U.H.E. (1997)
{\it J. Phys. Chem. B} {\bf 101}, 3304--3310.
\bibitem{HNSKN} Higo, J., Nakajima, N., Shirai, H., Kidera, A. \& Nakamura, H.
(1997) {\it J. Comput. Chem.} {\bf 18}, 2086--2092.
\bibitem{NakaHKN} Nakajima, N., Higo, Kidera, A. \& Nakamura, H.
(1997) {\it J. Chem. Phys.} {\bf 278}, 297--301.
\bibitem{NY} Noguchi, H. \& Yoshikawa, K. (1997)
{\it Chem. Phys. Lett.} {\bf 278}, 184--188.
\bibitem{KGS2} Kolinski, A., Galazka, W. \& Skolnick, J.
(1998) {\it J. Chem. Phys.} {\bf 108}, 2608--2617.
\bibitem{ICK} Iba, Y., Chikenji, G. \& Kikuchi, M. (1998) {\it J. Phys. Soc.
 Jpn.}
{\bf 67}, 3327--3330.
\bibitem{N} Nakajima, N. (1998) {\it Chem. Phys. Lett.} {\bf 288}, 319--326.
\bibitem{HSchp} Hao, M.H. \& Scheraga, H.A. (1998) {\it J. Mol. Biol.} 
{\bf 277}, 973--983.
\bibitem{SNHKN} Shirai, H., Nakajima, N., Higo, J., Kidera, A. \& Nakamura, H.
(1998) {\it J. Mol. Biol.} {\bf 278}, 481--496.

\bibitem{SBK} Schaefer, M., Bartels, C. \& Karplus, M. (1998)
{\it J. Mol. Biol.} {\bf 284}, 835--848.
\bibitem{MHO} Mitsutake, A., Hansmann, U.H.E. \& Okamoto, Y. (1998)
{\it J. Mol. Graphics Mod.} {\bf 16}, 226--238; 262--263.
\bibitem{HO99} Hansmann, U.H.E. \& Okamoto, Y. (1999)
{\it J. Phys. Chem. B} {\bf 103}, 1595--1604.
\bibitem{SUYH} Shimizu, H., Uehara, K., Yamamoto, K. \& Hiwatari, Y.
(1999) {\it Mol. Sim.} {\bf 22}, 285--301.
\bibitem{ONHN} Ono, S., Nakajima, N., Higo, J. \& Nakamura, H. (1999)
{\it Chem. Phys. Lett.} {\bf 312}, 247--254.
\bibitem{MO2} Mitsutake, A. \& Okamoto, Y. (2000)
{\it J. Chem. Phys.} {\bf 112}, 10638--10647.
\bibitem{YCBM} Yasar, F., Celik, T., Berg, B.A. \& Meirovitch, H. (2000)
{\it J. Comput. Chem.} {\bf 21}, 1251--1261.
\bibitem{MO3} Mitsutake, A., Kinoshita, M., Okamoto, Y. \& Hirata, F.
(2000) {\it Chem. Phys. Lett.} {\bf 329}, 295--303.
\bibitem{Muna} Munakata, T. \& Oyama, S. (1996) {\it Phys. Rev. E}
{\bf 54}, 4394--4398. 

\bibitem{ST1} Lyubartsev, A.P., Martinovski, A.A., Shevkunov, S.V. \&
Vorontsov-Velyaminov, P.N. (1992) {\it J. Chem. Phys.} {\bf 96}, 
1776--1783.
\bibitem{ST2} Marinari E. \& Parisi, G. (1992) {\it Europhys. Lett.} 
{\bf 19}, 451--458.
\bibitem{STrev} Marinari, E., Parisi, G. \& Ruiz-Lorenzo, J.J. (1998)
in {\it Spin Glasses and Random Fields}, Young, A.P., Ed.,
World Scientific, Singapore, pp. 59--98.
\bibitem{IRB1} Irb{\"a}ck, A. \& Potthast, F. (1995) {\it J. Chem. Phys.}
{\bf 103}, 10298--10305.
\bibitem{HO96a} Hansmann, U.H.E. \& Okamoto, Y. (1996) {\it Phys. Rev. E} 
{\bf 54}, 5863--5865.
\bibitem{HO96b} Hansmann, U.H.E. \& Okamoto, Y. (1997) {\it J. Comput. Chem.} 
{\bf 18}, 920--933.
\bibitem{IRB2} Irb{\"a}ck, A. \& Sandelin, E. (1999) {\it J. Chem. Phys.}
{\bf 110}, 12256--12262.
\bibitem{HS} Hesselbo, B. \& Stinchcombe,\ R.B. (1995) {\it Phys. Rev. Lett.} {\bf 74}, 2151--2155.

\bibitem{SmBr} Smith, G.R. \& Bruce, A.D. (1996)
{\it Phys. Rev. E} {\bf 53}, 6530--6543.
\bibitem{H97c} Hansmann, U.H.E. (1997)
{\it Phys. Rev. E} {\bf 56}, 6201--6203.
\bibitem{MUCAW} Berg, B.A. (1998)
{\it Nucl. Phys. B} (Proc. Suppl.) {\bf 63A-C}, 982--984.
\bibitem{Tsa} Tsallis, C. (1988) {\it J. Stat. Phys.}
{\bf 52}, 479--487.
\bibitem{HO96d} Hansmann, U.H.E. \& Okamoto, Y. (1997)
{\it Phys. Rev. E} {\bf 56}, 2228--2233.
\bibitem{HEO98} Hansmann, U.H.E., Eisenmenger, F. \& Okamoto, Y. (1998)
{\it Chem. Phys. Lett.} {\bf 297}, 374--382.
\bibitem{HMO97} Hansmann, U.H.E., Masuya, M. \& Okamoto, Y. (1997)
{\it Proc. Natl. Acad. Sci. U.S.A.} {\bf 94}, 10652--10656.
\bibitem{HOO} Hansmann, U.H.E., Okamoto, Y. \& Onuchic, J.N. (1999)
{\it Proteins} {\bf 34}, 472--483.
\bibitem{Str2} Andricioaei, I. \& Straub, J.E. (1997)
{\it J. Chem. Phys.} {\bf 107}, 9117--9124.
\bibitem{Muna2} Munakata, T. \& Mitsuoka, S. (2000) {\it J. Phys. Soc.
 Jpn.}
{\bf 69}, 92--96.
\bibitem{SA} Kirkpatrick, S., Gelatt, C.D. Jr. \& Vecchi, M.P.
   (1983) {\it Science} {\bf 220}, 671--680. 
\bibitem{STsal} Tsallis, C. \& Stariolo, D.A. (1996)
 {\it Physica} {\bf A233}, 395--406.
\bibitem{Str1} Andricioaei, I. \& Straub, J.E. (1996)
{\it Phys. Rev. E} {\bf 53}, R3055--R3058.
\bibitem{H97b} Hansmann, U.H.E. (1998)
{\it Physica A} {\bf 242}, 250--257.
\bibitem{BeSt} Berne, B.J. \& Straub, J.E. (1997)
{\it Curr. Opin. Struct. Biol.} {\bf 7}, 181--189.
\bibitem{RevStr} Straub, J.E. \& Andricioaei, I. (1999)
{\it Braz. J. Phys.} {\bf 29}, 179--186.
\bibitem{RevHO3} Hansmann, U.H.E. \& Okamoto, Y. (1999)
{\it Braz. J. Phys.} {\bf 29}, 187--198.
\bibitem{RE1} Hukushima, K. \&  Nemoto, K. (1996)
{\it J. Phys. Soc. Jpn.} {\bf 65}, 1604--1608.
\bibitem{RE1b} Hukushima, K., Takayama, H. \& Nemoto, K. (1996)
{\it Int. J. Mod. Phys. C} {\bf 7}, 337--344.

\bibitem{RE2} Geyer, C.J. (1991) in {\it Computing Science and Statistics:
 Proc. 23rd Symp. on the Interface}, Keramidas, E.M., Ed.,
 Interface Foundation, Fairfax Station, pp. 156--163.
\bibitem{RE3} Swendsen, R.H. \& Wang, J.-S. (1986)
 {\it Phys. Rev. Lett.} {\bf 57}, 2607--2609.
\bibitem{RE4} Tesi, M.C., van Rensburg, E.J.J., Orlandini, E. \&
Whittington, S.G. (1996) {\it J. Stat. Phys.} {\bf 82}, 155--181.

\bibitem{H97} Hansmann, U.H.E. (1997)
{\it Chem. Phys. Lett.} {\bf 281}, 140--150. 
\bibitem{FD} Falcioni, M. \& Deem, M.W. (1999)
{\it J. Chem. Phys.} {\bf 111}, 6625--6632. 

\bibitem{SO} Sugita, Y. \& Okamoto, Y. (1999)
{\it Chem. Phys. Lett.} {\bf 314}, 141--151.
\bibitem{Kol} Gront, D., Kolinski, A. \& Skolnick, J. (2000) 
{\it J. Chem. Phys.} {\bf 113}, 5065--5071.
\bibitem{SKO} Sugita, Y., Kitao, A. \& Okamoto, Y. (2000)
{\it J. Chem. Phys.} {\bf 113}, 6042--6051.
\bibitem{Gar} Garcia, A.E. \& Sanbonmatsu, K.Y. (2000)
\lq\lq Exploring the energy landscape of a $\beta$ hairpin in explicit 
solvent,\rq\rq~ {\it Proteins}, in press.

\bibitem{YP} Yan, Q. \& de Pablo, J.J. (1999)
{\it J. Chem. Phys.} {\bf 111}, 9509--9516. 

\bibitem{NOSMO} Nishikawa, Y., Ohtsuka, H., Sugita, Y., 
Mikami, M. \& Okamoto, Y. (2000)
{\it Prog. Theor. Phys. (Suppl.)} {\bf 138}, 270--271.
\bibitem{Freeman} Calvo, F., Neirotti, J.P., Freeman, D.L. \& Doll, J.D.
(2000) {\it J. Chem. Phys.} {\bf 112}, 10350--10357.
\bibitem{OKOM} Okabe, T., Kawata, M., Okamoto, Y. \& Mikami, M. (2000)
\lq \lq Replica-exchange Monte Carlo method for the isobaric-isothermal
ensemble,\rq\rq~ submitted for publication.
\bibitem{ISNO} Ishikawa, Y., Sugita, Y., Nishikawa, T. \& Okamoto, Y. (2000)
\lq\lq {\it Ab initio} replica-exchange Monte Carlo method for
cluster studies,\rq\rq~{\it Chem. Phys. Lett.}, in press.

\bibitem{Yama} Yamamoto, R. \& Kob, W. (2000)
{\it Phys. Rev. E} {\bf 61}, 5473--5476. 
\bibitem{Huk2} Hukushima, K. (1999)
{\it Phys. Rev. E} {\bf 60}, 3606--3614. 
\bibitem{Dunw} Bunker, A. \& D{\" u}nweg, B. (2000)
\lq\lq Parallel excluded volume tempering for polymer melts,\rq\rq~
{\it Phys. Rev. E}, in press.

\bibitem{SO3} Sugita, Y. \& Okamoto, Y. (2000)
{\it Chem. Phys. Lett.} {\bf 329}, 261--270.
\bibitem{MO4} Mitsutake, A., \& Okamoto, Y. (2000) 
{\it Chem. Phys. Lett.} {\bf 332}, 131--138.

\bibitem{Metro} Metropolis, N., Rosenbluth, A.W., Rosenbluth, M.N.,
Teller, A.H. \& Teller, E. (1953) {\it J. Chem. Phys.} {\bf 21},
1087--1092.

\bibitem{ECEP1} 
Momany, F.A., McGuire, R.F., Burgess, A.W. \& Scheraga, H.A.
(1975) {\it J. Phys. Chem.} {\bf 79}, 2361--2381.
\bibitem{ECEP2} 
N{\'e}methy, G., Pottle, M.S. \& Scheraga, H.A.  (1983) 
{\it J. Phys. Chem.} {\bf 87}, 1883--1887.
\bibitem{ECEP3} 
Sippl, M.J., N{\'e}methy, G. \& Scheraga, H.A.  (1984) 
{\it J. Phys. Chem.} {\bf 88}, 6231--6233.
\bibitem{KONF1} Kawai, H., Okamoto, Y., Fukugita, M., Nakazawa, T. \& 
Kikuchi, T. (1991)
{\it Chem. Lett.} {\bf 1991}, 213--216.
\bibitem{KONF2} Okamoto, Y., Fukugita, M., Nakazawa, T. \& Kawai, H.
(1991) {\it Protein Eng.} {\bf 4}, 639--647.
\bibitem{SIG1} Hingerty, B.E., Ritchie, R.H., Ferrell, T. \& Turner, J.E.
(1985) {\it Biopolymers} {\bf 24}, 427--439.
\bibitem{SIG2} Ramstein, J. \& Lavery, R.
(1988) {\it Proc. Natl. Acad. Sci. U.S.A.} {\bf 85}, 7231--7235.
\bibitem{O2} Okamoto, Y.
(1994) {\it Biopolymers} {\bf 34}, 529--539.
\bibitem{DKK} Daggett, V., Kollman, P.A. \& Kuntz, I.D.
(1991) {\it Biopolymers} {\bf 31}, 285--304.
\bibitem{OONS} Ooi, T., Oobatake, M., N{\'e}methy, G. \& Scheraga, H.A.  
(1987) 
{\it Proc. Natl. Acad. Sci. U.S.A.} {\bf 84}, 3086--3090.
\bibitem{SOL2} Masuya, M., in preparation.
\bibitem{SOL3} Eisenhaber, F., Lijnzaad, P., Argos, P., Sander, C.
\& Scharf, M. (1995) {\it J. Comput. Chem.} {\bf 16}, 273--284.
\bibitem{RISM1} Chandler, D. \& Andersen, H.C. (1972)
{\it J. Chem. Phys.} {\bf 57}, 1930--1937.
\bibitem{RISM2} Hirata, F. \& Rossky, P.J. (1981)
{\it Chem. Phys. Lett.} {\bf 83}, 329--334.
\bibitem{RISM3} Perkyns, J.S. \& Pettitt, B.M. (1992)
{\it J. Chem. Phys.} {\bf 97}, 7656--7666.
\bibitem{SPC} Berendsen, H.J.C., Grigera, J.R. \& Straatsma, T.P.
(1987) {\it J. Phys. Chem.} {\bf 91}, 6269--6271.
\bibitem{KINO} Kinoshita, M., Okamoto, Y. \& Hirata, F. (1997)
{\it J. Comput. Chem.} {\bf 18}, 1320--1326.
 
\bibitem{AMBER1} Weiner, S.J., Kollman, P.A., Nguyen, D.T. \& Case, D.A. (1986)
{\it J. Comput. Chem.} {\bf 7}, 230--252.
\bibitem{AMBER2} Cornell, W.D., Cieplak, P., Bayly, C.I., Gould, I.R.,
Merz, K.M. Jr., Ferguson, D.M., Splellmeyer, D.C., Fox, T.,
Caldwell, J.W. \& Kollman, P.A. (1995)
{\it J. Am. Chem. Soc.} {\bf 117}, 5179--5197.
\bibitem{AMBER3} Kollman, P., Dixon, R., Cornell, W., Fox, T.
Chipot, C. \& Pohorille, A. (1997)
in {\it Computer Simulation of Biomolecular Systems Vol. 3},
van Gunsteren, W.F., Weiner, P.K. \& Wilkinson, A.J., Eds.,
KLUWER/ESCOM, Dordrecht, pp. 83--96.

\bibitem{SK} Sugita Y. \& Kitao, A. (1998) {\it Proteins} {\bf 30}, 388--400.
\bibitem{KHG} Kitao, A., Hayward, S. \& G\={o}, N. (1998) {\it Proteins}
{\bf 33}, 496--517.
\bibitem{PRESTO} Morikami, K., Nakai, T., Kidera, A., Saito, M. \&
Nakamura, H. (1992) {\it Comput. Chem.} {\bf 16}, 243--248.
\bibitem{HLM} Hoover, W.G., Ladd, A.J.C. \& Moran, B. (1982) 
{\it Phys. Rev. Lett.} {\bf 48}, 1818--1820.
\bibitem{EM} Evans, D.J. \& Morris, G.P. (1983) 
{\it Phys. Lett.} {\bf A98}, 433--436.
\bibitem{TIP3P} Jorgensen, W.L., Chandreskhar, J., Madura, J.D.,
Impey, R.W. \& Klein, M.L. (1982)
{\it J. Chem. Phys.} {\bf 79}, 926--935.

\bibitem{Scholtz} Myers, J.K., Pace, C.N. \& Scholtz, J.M.
(1997)
{\it Proc. Natl. Acad. Sci. U.S.A.} {\bf 94}, 2833--2837.
\bibitem{MO5} Mitsutake, A. \& Okamoto, Y. (2000), in preparation.

\bibitem{Molscript} Kraulis, P. J. (1991) {\it J. Appl. Cryst.} 
{\bf 24}, 946--950.
\bibitem{Raster3Da} Bacon, D. \& Anderson, W. F. (1988)
{\it J. Mol. Graphics}, {\bf 6}, 219--220.
\bibitem{Raster3Db} 
Merritt, E. A. \& Murphy, M. E. P. (1994) {\it Acta Cryst.}
{\bf D50}, 869--873.

\bibitem{RasMol} 
Sayle, R.A. \& Milner-White, E.J. (1995)
{\it Trends Biochem. Sci.}
{\bf 20}, 374--376.

\bibitem{MSO} Mitsutake, A., Sugita, Y. \& Okamoto, Y.
(2000), in preparation.

\bibitem{EnkNMR} Graham, W.H., Carter, S.E. II \&
Hickes, P.R. (1992) {\it Biopolymers}
{\bf 32}, 1755--1764. 

\bibitem{SO4} Sugita, Y. \& Okamoto, Y.
(2000), in preparation.

\bibitem{Ono} Ono, S., Nakajima, N., Higo, J. \& Nakamura, H. (2000)
{\it J. Comput. Chem.} {\bf 21}, 748--762.
\bibitem{KINO2} Kinoshita, M., Okamoto, Y. \& Hirata, F. (2000)
{\it J. Am. Chem. Soc.} {\bf 122}, 2773--2779.




\end{thebibliography}
\end{document}